\documentclass[12pt]{article}

\usepackage{graphicx,amsmath,amsfonts,amssymb,amsthm, amscd}

\def\ov{\overline}
\def\wh{\widehat}
\def\wt{\widetilde}
\def\pa{\partial}
\def\op{\operatorname}
\def\a{\alpha}
\def\be{\beta}
\def\ga{\gamma}
\def\dt{\delta}
\def\e{\varepsilon}
\def\th{\theta}
\def\ka{\varkappa}
\def\la{\lambda}
\def\si{\sigma}
\def\ph{\varphi}
\def\om{\omega}
\def\Ga{\Gamma}
\def\De{\Delta}
\def\Th{\Theta}
\def\Om{\Omega}
\def\La{\Lambda}
\def\N{\mathbb{N}}
\def\Z{\mathbb{Z}}
\def\R{\mathbb{R}}
\def\C{\mathbb{C}}
\renewcommand{\Im}{\operatorname{Im}}
\renewcommand{\Re}{\operatorname{Re}}

\newtheorem{Theorem}{Theorem}[section]

\newtheorem{RemarK}[Theorem]{Remark}
\newtheorem{LemmA}[Theorem]{Lemma}

\newtheorem{PropositioN}[Theorem]{Proposition}
\newtheorem{DefinitioN}[Theorem]{Definition}
\newtheorem{ExamplE}[Theorem]{Example}

\begin{document}

\title{The Complex Lagrangian Germ\\
and the Canonical Operator}

\author{V.~L.~Dubnov\\
{\small Moscow Institute of Electronics and Mathematics, Russia}
\\
V.~P.~Maslov\\
{\small National Research University ``Higher School of Economics'',}
\\
{\small Moscow Institute of Electronics and Mathematics, Russia}
\\
V.~E.~Nazaikinskii
\\
{\small Moscow Institute of Electronics and Mathematics, Russia}
}

\date{}

\maketitle

\begin{abstract}
We give a manifestly invariant definition of the Lagrangian complex germ with
the minimal degree of accuracy required to define the canonical operator. The
equivalence with the traditional definition is proved, and the canonical
operator is constructed in new terms.  A new form of the quantization
condition  is given, in which the volume form is assumed to be defined on
the universal covering of the Lagrangian manifold rather than on the
manifold itself. This allows one to solve a wider class of eigenvalue
problems.
\end{abstract}

\setcounter{section}{-1}

\section{Introduction}

The complex WKB method, originally developed in \cite{M1, M2}, is a
method for constructing asymptotic solutions to $1/h$-(pseudo)differential
equations. In the simplest case, it deals with asymptotic expansions of
the form
\begin{equation}
\psi(x,h)= \exp\Big\{\frac{i}{h} S(x)\Big\}\sum h^k \ph_k(x) + O(h^N),
\tag{0.1}
\end{equation}
where $S$ and $\ph_k$ are smooth functions and $\Im S\geqslant0$.

The following observation is of crucial importance:
$$
\ph(x)\exp\Big\{\frac{i}{h} S(x)\Big\}=O(h^s)
$$
whenever $\ph(x)=O((\Im S)^s)$, and $\pa\Im S/\pa x=O((\Im S)^{1/2})$.
Thus, expansions in powers of $\Im S$ and $\pa\Im S/\pa x$ in the
amplitude result in expansions in powers of $h$ in \thetag{0.1}, and the
Hamilton-Jacobi equation $H(x,\pa S/\pa x)=0$ makes sense and can be
solved even for nonanalytic Hamiltonians, since it can be understood as an
asymptotic expansion in powers of $\pa(\Im S)/\pa x$. It turns out that
the minimum accuracy in specifying $S(x)$ which still allows one to obtain
asymptotic expansions of solutions modulo $O(h^N)$ with arbitrary $N$ is
$O((\Im S)^{3/2})$. Accordingly, the accuracy in specifying the associated
Lagrangian manifold
$L:\{(x,p)\mid p={\pa S}/{\pa x}\}$ is $O(\Im S)$. This {\it minimum
accuracy condition\/} was followed in \cite{M2}, but an unnecessary
assumption was made, namely, that there is an underlying real Lagrangian
manifold invariant with respect to the real part of the Hamiltonian vector
field.

For this reason the results of \cite{M2} apply mainly to the Cauchy problem
and cannot be used directly in solving general eigenvalue problems.

Various subsequent expositions of the complex germ theory (another name
for the complex WKB method) either fail to satisfy the minimum accuracy
condition \cite{Ku1, Ku2, MeS1, MeS2, MSS1, MSS2}, or deal only with the
special case of the complex germ over an isotropic manifold ($\Im S_2=0$ on
some submanifold and $\Im S_2$ has a nondegenerate Hessian in the
transversal direction) \cite{BD1, BD2, M3, Vo}, or fail to provide an
invariant geometric description \cite{VDM1, VDM2}.

It was proposed in \cite{H} to describe the Lagrangian manifold $L$ by
using the ideal generated by $p_i-\pa S/\pa x_i$, $i=1,\dots,n$, in the
spirit of algebraic geometry. However, the impact of the damping factor
$\exp\{-\Im S/h\}$ on the geometry was not investigated in \cite{H},
and hence the canonical operator cannot be constructed on the basis of
these results.

Here we give a definition of Lagrangian asymptotic manifolds resembling
that in \cite{VDM2} but incorporating the minimum accuracy condition. We
prove its equivalence to the traditional definition and construct the
canonical operator in these new terms. A new form of the quantization
condition  is given, in which the volume form is assumed to be defined on
the universal covering of the Lagrangian manifold rather than on the
manifold itself. This allows one to solve a wider class of eigenvalue
problems.

We also suggest a new approach of defining the
dissipation in the canonical charts as the minimum of a dissipation
globally defined on the phase space. The canonical operator to the
accuracy of $O(h^\infty)$ is defined in \S5 with the help of the so-called
$V$-{\it objects\/}, originally introduced in the real case in \cite{M2}
and closely related to the famous Atiyah group operators.

The first four sections have been written by V.~P.~Maslov and
V.~E.~Nazaikinskii, and the fifth section by V.~P.~Maslov and V.~L.~Dubnov.

\section{Asymptotic manifolds}

Throughout this section, $M$ will be an $n$-dimensional  differential real
manifold. By $\mathcal{C}^\infty(M)$ we denote the sheaf of germs of
{\it complex-valued\/} $C^\infty$ functions on $M$. Unless otherwise specified,
vector fields, differential forms, etc., are allowed to have complex-valued
coefficients.

We are not too pedantic about the distinction between sections  and
elements of sheaves. Occasionally, we may write something like
$f\in \mathcal{C}^\infty(M)$ instead of the formally correct
$f\in \Ga(U,\mathcal{C}^\infty(M))$.
However, such liberties are generally harmless (since the
sheaves we consider are fine) and the set $U$ is always obvious from the
context.

\subsection{Dissipations and dissipation ideals}

\begin{DefinitioN}\rm
A {\it dissipation\/} on $M$ is a smooth
nonnegative function $D:M\to\R$. Two dissipations $D_1$ and $D_2$ are said to
be {\it equivalent\/}, $D_1\sim D_2$, if locally (in the vicinity of any point
$m_0\in M$) we have
$$
c_1D_1(m)\leqslant D_2(m)\leqslant c_2D_2(m)
$$
with some positive constants $c_1$ and $c_2$.
\end{DefinitioN}

\begin{DefinitioN} 
\rm
Let $D$ be a dissipation on $M$. Consider the
sheaf of ideals $\mathcal{D}\subset\mathcal{C}^\infty(M)$ such that for any $m\in M$ the
stalk $\mathcal{D}_m$ is the set of germs  $f\in\mathcal{C}_m^\infty(M)$ of
functions $\wt f$ satisfying the estimate
$$
|\wt f|\leqslant cD
$$
with some constant $c\geqslant 0$. The sheaf $\mathcal{D}$ is called the {\it
dissipation ideal\/} associated with $D$.
\end{DefinitioN}

Obviously, a dissipation ideal depends only on the equivalence class of the
corresponding dissipation, rather than on the dissipation itself.

Let $\mathcal{D}$ be a dissipation ideal on $M$. The locus $\op{loc}(\mathcal{D})$
(i.e., the set of common zeros of all sections of $\mathcal{D}$) will be denoted by
$\Ga$ (or by $\Ga_{\mathcal{D}}$ if there is any risk of confusion). Equivalently,
$\Ga$ can be characterized as the support
of the quotient sheaf $\mathcal{C}^\infty(M)/\mathcal{D}$. It is obvious that
$$
\Ga = \{m\in M\mid D(m)=0\}
$$
for  any dissipation $D$ associated with $\mathcal{D}$.

Given a dissipation ideal $\mathcal{D}\subset \mathcal{C}^\infty(M)$, for any
$s\geqslant 0$ we construct an ideal ${\mathcal{D}}^s$
as follows. Let $D$ be some dissipation associated with $\mathcal{D}$. We define
the stalk $\mathcal{D}_m^s$ to be the set of germs  $f\in \mathcal{C}^\infty_m(M)$
of functions $\wt f$ satisfying the estimate
$$
|\wt f|\leqslant c D^s
$$
for some constant $c\geqslant 0$. Obviously, $\mathcal{D}^s\supset{\mathcal{D}}^k$
for $s\leqslant k$; furthermore, $\mathcal{D}^s \mathcal{D}^k\subset \mathcal{D}^{s+k}$,
but the inclusion is strict in general, i.e.,
$\mathcal{D}^s \mathcal{D}^k\ne \mathcal{D}^{s+k}$.
Although the definition of $\mathcal{D}^s$ makes sense for any
$s\geqslant 0$, we mainly use the ideals $\mathcal{D}^s$ for $s=N/2$, where
$N$ is a positive integer. The reason for introducing half-integer values of
$s$ in our considerations is clear from the following lemma.

\begin{LemmA} 
Let $\mathcal{D}$ be a dissipation ideal on $M$. If $f\in \mathcal{D}^s$ for some
$s\geqslant 1/2$  and if $X$ is a smooth vector field on $M$, then $Xf\in\mathcal{D}^{s-1/2}$.
\end{LemmA}

\noindent{\it Proof}.
The statement of the lemma is equivalent to saying that
$$
\Big|\frac{\pa^{|\a|}f(x)}{\pa x^\a}\Big|\leqslant c_\a(D(x))^{s-|\a|/2},\quad
|\a|\leqslant 2s,
$$
whenever $|f(x)|\leqslant c(D(x))^s$ (here $c$ and $c_\a$ are constants). The
proof of the latter statement can be found in \cite{VDM1}, pp.~20--23.
\qed
\medskip

The following ideals are sometimes useful, as well as ${\mathcal{D}}^s$. Set
$$
\overset{\circ}{\mathcal{D}}^s=\bigcap_{\e>0} {\mathcal{D}}^{s-\e},\quad s>0,\quad
\overset{\circ}{\mathcal{D}}\equiv \overset{\circ}{\mathcal{D}}^1.
$$
Obviously, for any $k,s\geqslant 0$ with $s>k$, we have ${\mathcal{D}}^s\subset
\overset{\circ}{\mathcal{D}}^s\subset {\mathcal{D}}^k$, and if $X$ is a smooth
vector field on $M$, then
$X\overset\circ{\mathcal{D}}^s\subset \overset{\circ}{\mathcal{D}}^{s-1/2}$
for any $s>1/2$. It can happen that not all vector fields behave
that badly. Very frequently we shall use vector fields for which the ideals
$\overset{\circ}{\mathcal{D}}^s$ are invariant, that is,
$X\overset{\circ}{\mathcal{D}}^s\subset \overset{\circ}{\mathcal{D}}^s$ for any $s$.

Let $\mathcal{D}$ be a dissipation ideal on $M$, let $D$ be some dissipation
associated with $\mathcal{D}$, and let $X$ be a smooth vector field on $M$.

\begin{LemmA} 
{\rm i)} Suppose that $XD\in\overset{\circ}{\mathcal{D}}$.
Then $X\overset{\circ}{\mathcal{D}}^s\subset \overset{\circ}{\mathcal{D}}^s$ for any $s>0$.

{\rm ii)} The inclusion $XD\in\mathcal{D}$ (and even $XD=0$) does not imply
$X{\mathcal{D}}^s\subset{\mathcal{D}}^s$ in general.
\end{LemmA}

Needless to say, it is item ii) that prompts the introduction of the ideals
$\overset{\circ}{\mathcal{D}}^s$. In many cases (say, for linear
complex germs \cite{M3}, or, more generally, if the dissipation is
real-analytic) we have $\overset{\circ}{\mathcal{D}}^s={\mathcal{D}}^s$ and thus can
avoid all these complications.
\medskip

\noindent
{\it Proof of Lemma~{\rm1.4}}.
(i) It suffices to prove that for any
positive integer $r\geqslant 2$ and any $s\geqslant 1/r$ we have
\begin{equation}
X{\mathcal{D}}^s\subset{\mathcal{D}}^{s-1/r}.
\tag{1.1}
\end{equation}
For $r=2$, assertion \thetag{1.1} is valid, since it is just the statement
of Lemma~1.3. Let us carry out the induction step, that is, assume
\thetag{1.1} and prove that then $X{\mathcal{D}}^s\subset {\mathcal{D}}^{s-(1/r+1)}$.

Let $g^t$ be the (possibly local) phase flow corresponding to the vector
field $X$ (we assume that $X$ is real; if this is not the case, we simply
prove the lemma for $\Re X$ and $\Im X$ separately). By Taylor's formula,
for sufficiently small $t$
\begin{align*}
D(g^{tD^{1/(r+1)}(m)}(m)) &= D(m) + tD^{1/(r+1)}(m)(XD)(m)
\\
&\quad+ \sum_{k=2}^r \frac{t^k}{k!}D^{k/(r+1)}(m) X^{k-1}XD(m)+O (t^{r+1}D(m)).
\end{align*}
Let us estimate each term on the right-hand side in this equation. The
first term (we omit the argument $m$) is $O(D)$, and
the second term is $O(D^{1/(r+1)}D^{1-\e})=O(D)$ by the hypothesis of the
lemma. Next, we have
$X^{k-1}(XD) = O\big(D^{1-\e-(k-1)/r}\big)$
by the inductive assumption (note that $1-(k-1)/r>0$ for $k\leqslant r$
and that $\e$ is arbitrarily small); hence, each term in the sum is
$$
O\big(D^{k/(r+1)+1-\e-(k-1)/r}\big) = O\big(D^{1-\e+(r-k+1)/(r(r+1))}\big)
= O(D)
$$
since $k\leqslant r$ and $\e$ is arbitrarily small. We conclude that
$D(g^{tD^{1/(r+1)}(m)}(m)) = O(D(m))$.

Now let $f\in{\mathcal{D}}^s$. Again by Taylor's formula,
$$
f(g^{tD^{1/(r+1)}(m)}(m)) = \sum_{k=0}^{N-1}\frac{t^k}{k!}D^{k/(r+1)}(m)
(X^kf)(m) + O(D(m)^{N/(r+1)}).
$$
Let us take $N$ to be the least positive integer such that
$N/(r+1)\geqslant s$. By the preceding, the left-hand side in the last
equation is
$O(D^s(g^{tD^{1/(r+1)}(m)}(m)) = O(D^s(m))$,
and we see that
$$
\sum_{k=0}^{N-1}\frac{t^k}{k!}D^{k/(r+1)}(m)(X^k f)(m)=O(D^s(m))
$$
for any sufficiently small $t$. Let us consider the last equation for $N$
distinct values $t=t_1,\dots, t=t_N$. Since the Vandermonde determinant
$$
\left|
\begin{matrix}
 1 & t_1 &\dots &t_1^{N-1}\\
\dots &\dots &\dots &\dots\\
1 &t_N &\dots &t_N^{N-1}
\end{matrix}
\right|
$$
is nonzero, it follows that $D^{k/(r+1)} X^k f= O(D^s)$; in particular,
for $k=1$, we obtain $Xf=O(D^{s-1/(r+1)})$, as desired.

ii) Consider the following example:
\begin{align*}
M&= \R^2\ni x=(x_1,x_2); \quad D(x) = \exp(-1/x_2^2);\quad
X= \frac{\pa}{\pa x_1};\\
f(x)&= D(x) \sin(x_1/x_2)\quad (D(x)=f(x)=0\,\,\text{ for }\,\,x_2=0).
\end{align*}
Then $XD=0$ and $f=O(D)$, but
$Xf=\frac{1}{x_2}\cos(x_1/x_2) D(x) \ne O(D)$.
The lemma is proved.
\qed \medskip

If $X$ is a vector field satisfying the conclusion of Lemma~1.4 i), then
we say that the dissipation ideal is  {\it invariant with respect to\/}
$X$.

\begin{DefinitioN} \rm
Let $m_0\in\Ga$, and let $U\subset M$ be a neighborhood of $m_0$. A smooth
mapping
$\ph:U\to M$
is called an {\it almost-identity diffeomorphism\/} if $\ph$ is a
diffeomorphism of $U$ onto $\ph(U)$ and
$\op{dist}(\ph(m),m)\leqslant c(D(m))^{1/2}$, $m\in U$,
where $D$ is some dissipation associated with $\mathcal{D}$, $c>0$, and
$\op{dist}(\cdot,\cdot)$ is the distance function induced by some Riemannian
metric on $M$.
\end{DefinitioN}
\medskip

\noindent{\bf Remark}.
Note that $\ph(m)=m$ for any $m\in\Ga$.
\medskip

\begin{LemmA} 
Let $\ph$ be an almost-identity diffeomorphism near $m_0\in\Ga$. Then
$\ph^{-1}$is also an almost-identity diffeomorphism, each of the ideals $\mathcal{D}^s$
is invariant by $\ph$, i.e.,
$\ph^*(\mathcal{D}^s)=\mathcal{D}^s$,
and if $\psi$ is another almost-identity diffeomorphism near $m_0$, then so is
$\ph\circ \psi$.
\end{LemmA}

\noindent{\it Proof} (see \cite{MN, VDM1}). We use local coordinates
$(x_1,\dots,x_n)$ on $M$ near $m_0$. By Ha\-damard's lemma,
$$
D(\ph(x)) = D(x) + \left\langle
\frac{\pa D(x)}{\pa x}, \ph(x)-x
\right\rangle + \left\langle \ph(x)-x, B(x)(\ph(x)-x)\right\rangle,
$$
where $B(x)$ is a smooth matrix function and
$\langle z,w\rangle= z_1 w_1 +\cdots +z_n w_n$.
By the assumption of the lemma,
$\|\ph(x)-x\|\leqslant c D(x)^{1/2}$,
and by Lemma~1.3
$$
\Big|\frac{\pa D(x)}{\pa x}\Big|\leqslant c_1 D(x)^{1/2}.
$$
Substituting the last two estimates into \thetag{1.1} yields
$$
D(\ph(x))\leqslant c_2 D(x).
$$
To prove the reverse inequality, let us apply Hadamard's lemma to the identity
$x=\ph(\ph^{-1}(x))$. We obtain
\begin{equation}
x-\ph(x) = A(x)(\ph^{-1}(x)-x),
\tag{1.2}
\end{equation}
where $A(x)$ is a smooth matrix function and $A(x_0)=(\pa\ph/\pa x) (x_0)$
(here $x_0$ is the coordinate image of $m_0$), since $\ph(x_0)=x_0$.
Consequently, $\op{det} A(x)\ne 0$ for $x$ close to $x_0$, and we obtain the
estimate
$$
\|\ph^{-1}(x)-x\|\leqslant c_3 D(x)^{1/2}
$$
by multiplying both sides in \thetag{1.2} by $A(x)^{-1}$. Hence, $\ph^{-1}(x)$
is an almost-identity diffeomorphism, and the inequality
$$
D(x)\leqslant c_4 D(\ph(x))
$$
follows from the above reasoning by symmetry.

The statement concerning the composition $\ph\circ\psi$ is trivial, and so the
lemma is proved.
\qed \medskip

In what follows we shall sometimes  use the notion of ``complex coordinates"
on the manifold $M$ (cf. \cite{MN, VDM1}). Suppose that $F_1,\dots,F_n\in
C^\infty(M)$, $n=\op{dim}M$, are smooth complex-valued functions on $M$. We
say that $F_1,\dots, F_n$ form a complex coordinate system in a
neighborhood of a point $x_0\in M$ if the differentials $dF_1,\dots,dF_n$
are linearly independent at $x_0$. Then for any function $f\in C^\infty(M)$ we have
$$
df=a_1\,dF_1+\cdots+ a_n\,dF_n
$$
near $x_0$, where the functions $a_1,\dots,a_n\in C^\infty(M)$ are
uniquely determined. They are referred to as the {\it partial
derivatives\/} of $f$ with respect to $F_1,\dots,F_n$ and denoted by
$$
a_j=\frac{\pa f}{\pa F_j}, \quad j=1,\dots,n;
$$
this coincides with the usual definition if $F_1=x_1,\dots, F_n=x_n$ is a
usual coordinate system on $M$. The derivatives thus defined retain such
familiar properties as
$$
\frac{\pa^2 f}{\pa F_j\pa F_j}=\frac{\pa^2 f}{\pa F_j\pa F_i}
$$
(this follows readily from the identity $d^2=0$) and
$$
\frac{\pa f}{\pa F_j} =\sum_{k=1}^n \frac{\pa f}{\pa Q_k}\frac{\pa Q_k}{\pa
F_j}
$$
for any coordinate system $Q_1,\dots,Q_n$ (complex or real).
One can often safely think of a function $f\in C^\infty(M)$
as $f=f(F_1,\dots,F_n)$, where $F_1,\dots,F_n$ are complex coordinates; if
a dissipation ideal $\mathcal{D}$ is given on $\mathcal{M}$ and there is a system of
functions $\De F_1,\dots,\De F_n \in{\mathcal{D}}^{1/2}$, then the {\it
asymptotic substitution\/} operator is defined
$$
(\si_{F\to F+\De F}^{(N)} f)(x) \simeq \sum_{|\a|=0}^N \frac{(\De
F)^\a}{\a!}\frac{\pa^\a f}{\pa F^\a},\qquad \a=(\a_1,\dots,\a_n)\in\Z_+^n.
$$
This operator represents the $N$th partial sum of the formal Taylor series
for $f(F_1+\De F_1,\dots, F_n+\De F_n)$ and possesses many
elegant properties for which we refer the reader to \cite{MN}, \cite{MSS1},
and \cite{VDM1}. Here we only note that
$$
\si_{F\to F+\De F}^{(N+l)}(f)-
\si_{F\to F+\De F}^{(N)}(f) = O(D^{(N+1)/2})
$$
for $l>0$ and that many of
our formulas arise from the specialization of the asymptotic substitution
operator to $N=1$ or $N=2$.

\subsection{Asymptotic submanifolds. Global definition}

The only kind of asymptotic manifolds used in this paper are asymptotic
submanifolds, and so it is in this case that we give detailed definitions,
lemmas, theorems, etc. However, it may be instructive at least to
sketch the intrinsic definitions.

Let $M$ be an $n$-dimensional manifold, and let $\mathcal{D}$ be a dissipation
ideal on $M$. Take some $s\geqslant 1$. The sheaf
$$
\mathcal{A}=(C^\infty(M)/{\mathcal{D}}^s)\big|_{\Ga_{\mathcal{D}}}
$$
is a sheaf of local rings on $\Ga= \Ga_{\mathcal{D}}$. Thus, $(\Ga,\mathcal{A})$ is
a ringed space, which will serve as a local model in our definition.
Namely, an $s$-asymptotic manifold of dimension $n$ is a ringed space
$(\mathcal{T}, \mathcal{B})$, where $\mathcal{T}$ is a Hausdorff locally compact space
and $\mathcal{B}$ is a sheaf of local rings on $\mathcal{T}$ such that for any
point $\ga_0\in\mathcal{T}$ there exist a neighborhood $U(\ga_0)\in\Ga$, a
dissipation ideal $\mathcal{D}$ on $\R^n$, and a mapping
$\tau:U(\ga_0)\to\R^n$ such that
\begin{enumerate}
\item[(a)] there is a neighborhood $U\subset \R^n$ of the set
$\tau(U(\ga_0))$ such that $U\cap\Ga_{\mathcal{D}}=\tau(U(\ga_0))$;
\item[(b)] $\tau$ is a homeomorphism of $U(\ga_0)$ into $U\cap\Ga_{\mathcal{D}}$
(the latter set is equipped with the topology inherited from $\R^n$);
\item[(c)] there is an isomorphism of sheaves
$\mathcal{B}\big|_{U(\ga_0)}\simeq \ga_0^{-1}({\mathcal{C}}^\infty(M)/{\mathcal{D}}^s)$.
\end{enumerate}
Then the construction goes along standard lines, under various auxiliary
assumptions (cf. \cite{VDM1}). However, the approach adopted here is more
straightforward; namely, we directly proceed to asymptotic submanifolds.

\begin{DefinitioN} 
Let $s\geqslant 1$. An $s$-{\it asymptotic
submanifold of codimension\/} $k$ {\it in\/} $M$ is a pair
$L=(\mathcal{D}, \mathcal{J})$, where $\mathcal{D}$ is a dissipation ideal
on $M$ and $\mathcal{J}\subset \mathcal{C}^\infty(M)$ is an ideal such that
\begin{enumerate}
\item[{\rm(i)}]
$\hfil\hfill\mathcal{D}^s\subset\mathcal{J}\subset \mathcal{D}^{1/2};\hfil\hfill (1.3)$
\item[{\rm(ii)}] in a neighborhood of each point
$m_0\in\Ga=\Ga_{\mathcal{D}}$ the ideal $\mathcal{J}$ is generated by $\mathcal{D}^s$ and by
$k$ functions $f_1,\dots,f_k$ such that the differentials $df_1,\dots,df_k$
are linearly independent at $m_0$ (in this case we say that $\mathcal{J}$ is $\mathcal{D}^s$-{\it nondegenerate of rank\/} $k$, and $f_1,\dots,f_k$ are referred to
as {\it generators modulo\/} $\mathcal{D}^s$ or simply {\it generators of\/}
$\mathcal{J}$).
\end{enumerate}
\end{DefinitioN}

Obviously, the number $k$ in condition (ii) is independent of the choice of
generators. Indeed, if $g_1,\dots,g_l$ is another system of generators, then
$$
g_j=\sum_{r=1}^k a_{jr}f_r+\ph_j,\quad j=1,\dots,l,
$$
where $a_{jr}$ are smooth functions and $\ph_j\in\mathcal{D}^s$. Since
$f_1(m_0)=\dots= f_k(m_0)=g_1(m_0)=\dots= g_l(m_0)=0$
and $d\ph_j(m_0)=0$, it follows that
$$
dg_j(m_0)=\sum_{r=1}^k a_{jr}(m_0)\,df_r(m_0),\quad j=1,\dots,l,
$$
and hence $l\leqslant k$ (the differentials $dg_j$ are assumed to be
linearly independent). By symmetry, $k\leqslant l$, and so  in fact
$k=l$.

As usual, the {\it dimension\/} of $L$ is defined by the formula
$\op{dim}L=\op{dim} M-k$.

\begin{ExamplE} 
\rm
Let $L\subset M$ be an ordinary submanifold of codimension $k$. Let us show
that it can be interpreted naturally as an $s$-asymptotic submanifold for
any $s$.

We introduce a dissipation on $M$ by setting
\begin{equation}
D(x)=(\op{dist}(x,L))^2,
\tag{1.4}
\end{equation}
where $\op{dist}(\cdot\,,\cdot)$ is the distance in some metric on $M$ (the
function \thetag{1.4} should be smooth; the metric can always be chosen so as
to satisfy this condition). Next, we set
$$
\mathcal{J}=\mathcal{J}_L=\{f\in\mathcal{C}^\infty(M)\mid f\big|_L\equiv 0\};
$$
that is, $\mathcal{J}$ is the {\it defining ideal\/} of $L$.

It is easy to verify that $\mathcal{J}=\mathcal{D}^{1/2}$ and that condition (ii) of
Definition~1.7 is also satisfied.

In what follows we chiefly use $1$-asymptotic submanifolds and refer to them
simply as asymptotic submanifolds.
\end{ExamplE}

\subsection{Asymptotic submanifolds. Nonparametric local description}

If $L\subset M$ is a $k$-codimensional submanifold that is diffeomorphically
projected on the coordinate plane $(x_{k+1},\dots,x_n)$ in a coordinate system
$(x_1,\dots, x_n)$ about a point $m_0\in L$, then locally (in a neighborhood
of $m_0$), $L$ can be described by equations of the form
\begin{equation}
x_1 = g_1(x_{k+1},\dots,x_n),
\dots,\,\,
x_k = g_k(x_{k+1},\dots,x_n),
\tag{1.5}
\end{equation}
where $g_1,\dots,g_k$ are smooth functions. Note that the defining ideal $\mathcal{J}_L$
is generated by the functions
$x_1-g_1(x_{k+1},\dots,x_n),\dots, x_k-g_k(x_{k+1},\dots,x_n)$
in this case. A description similar to \thetag{1.5} holds for asymptotic
submanifolds. Let us study this description in detail. Since our
considerations are purely local, we can assume that $M=\R^n$ with the
coordinates $x=(x', x'')$, where $x'=(x_1,\dots,x_k)$,
$x''=(x_{k+1},\dots,x_n)$.

We obtain a local description of an asymptotic submanifold by allowing
the functions $g_1,\dots,g_k$ in \thetag{1.5} to take complex values. More
precisely, let $d(x'')$ be a dissipation on $\R_{x''}^{n-k}$, and let
$g_1(x''),\dots,g_k(x'')$ be smooth complex-valued functions such that
\begin{equation}
|\Im g_j(x'')|\leqslant c d(x'')^{1/2},\quad j=1,\dots,k.
\tag{1.6}
\end{equation}
Set
\begin{equation}
D(x)=d(x'')+\sum_{j=1}^k|x_j-g_j(x'')|^2
\tag{1.7}
\end{equation}
and let $\mathcal{J}\subset \mathcal{C}^\infty(M)$ be the ideal generated by $\mathcal{D}$
and by the $k$ functions $x_1-g_1(x''),\dots,x_k-g_k(x'')$. Then the pair
$(\mathcal{D}, \mathcal{J})$ is an asymptotic submanifold in $M$, since all requirements
in Definition~1.7 are obviously satisfied.

\begin{RemarK} 
\rm
If we replace $d(x'')$ by an equivalent dissipation $\wt d(x'')$ and take
smooth functions $\wt g_1(x''),\dots,\wt g_k(x'')$ such that
$$
\wt g_j(x'')- g_j(x'') = O(d(x'')),\quad j=1,\dots,k,
$$
then the new data $(\wt d, \wt g_1,\dots, \wt g_k)$ define the same
asymptotic submanifold $(\mathcal{D}, \mathcal{J})$.
\end{RemarK}

Let us now proceed to the inverse problem.

\begin{Theorem} 1.10
{\bf(Implicit Function Theorem for asymptotic submanifolds).}
Let $L=(\mathcal{D}, \mathcal{J})$ be an asymptotic $k$-codimensional submanifold in
$M$, let $x_0\in\Ga$, and suppose that for some system of generators
$f_1(x),\dots,f_k(x)$ of the ideal $\mathcal{J}$, the condition
\begin{equation}
\op{det}\frac{\pa (f_1(x),\dots,f_k(x))}{\pa(x_1,\dots,x_k)}\ne 0
\tag{1.8}
\end{equation}
is satisfied at the point $x_0$. Then
\begin{enumerate}
\item[{\rm(a)}] the same condition is satisfied for any system of generators of
$\mathcal{J}$;
\item[{\rm(b)}] there exist a dissipation $d(x'')$ defined in a neighborhood of
$x''_0$ and smooth functions $g_1(x''),\dots, g_k(x'')$ such that
$$
\wt D(x) = d(x'') + \sum_{j=1}^k|x_j-g_j(x'')|^2
$$
is a dissipation associated with $\mathcal{D}$,
$$
|\Im g_j(x'')|\leqslant cd(x'')^{1/2},\quad j=1,\dots,k,
$$
and $\mathcal{J}$ is generated by $\mathcal{D}$ and by the functions
$$
x_1-g_1(x''),\dots, x_k-g_k(x'')
$$
in a neighborhood of $x_0$;
\item[{\rm(c)}] the dissipation $d(x'')$ is determined uniquely modulo the
equivalence relation introduced in Definition~{\rm1.1}, and the functions
$g_j(x'')$ are determined uniquely modulo $O(d(x''))$;
\item[{\rm(d)}] any function $\Phi(x)\in C^\infty(M)$ can be represented in
the following form in a neighborhood of $x_0$:
$$
\Phi(x) = \ph(x'')+\eta(x),
$$
where $\eta(x)\in\mathcal{J}$; the function $\ph(x'')$ is uniquely determined
modulo $O(d(x''))$.
\end{enumerate}
\end{Theorem}
\medskip

\noindent{\it Proof}.
(a) This assertion is obvious.

(b) Let $D(x)$ be a dissipation associated with $\mathcal{D}$ in a
neighborhood of $x_0$. First, let us prove that
\begin{equation}
\op{det}\frac{\pa^2 D}{\pa x' \pa x'}(x_0) \ne 0.
\tag{1.9}
\end{equation}
To this end, consider the function
$F(x) = |f_1(x)|^2+\cdots+|f_k(x)|^2$.
This function has the following properties:
\begin{gather}
0\leqslant F(x)\leqslant cD(x); \quad F(x_0)=D(x_0)=0;
\nonumber\\
\frac{\pa^2 F}{\pa x'_j \pa x'_l}(x_0) = \sum_{r=1}^k\Big(\frac{\pa
f_r(x_0)}{\pa x_j}\frac{\pa
\ov f_r(x_0)}{\pa x_j} + \frac{\pa
\ov f_r(x)}{\pa x_j}\frac{\pa
f_r(x_0)}{\pa x_l}
\Big)
\tag{1.10}
\end{gather}
(the bar denotes complex conjugation). It follows from the last equation in
\thetag{1.10} that for any  $\xi\in\R^k$ we have
\begin{equation}
\Big(\xi,\frac{\pa^2F}{\pa x' \pa x'}(x_0)\xi\Big) = 2\sum_{r=1}^k
\Big|\Big(\frac{\pa f_r}{\pa x'}(x_0)\xi\Big)\Big|^2\geqslant c_1|\xi|^2
\tag{1.11}
\end{equation}
since the vectors $(\pa f_r/\pa x')(x_0)$ form a basis in $\C^k$.
Furthermore, it follows from the first two equations in \thetag{1.10} that
$$
\frac{\pa^2D}{\pa x'\pa x'}(x_0)\geqslant \frac{1}{c}
\frac{\pa^2F}{\pa x'\pa x'}(x_0).
$$
By \thetag{1.11}, the matrix $(\pa^2F/\pa x' \pa x')(x_0)$ is positive
definite, and so, {\it a fortiori\/}, is the matrix
$(\pa^2D/\pa x' \pa x')(x_0)$.
In particular,
$\op{det}(\pa^2D/\pa x' \pa x')(x_0)\ne 0$.

Let us now consider the  equation
\begin{equation}
\frac{\pa D}{\pa x'}(x',x'')=0.
\tag{1.12}
\end{equation}
We have $(\pa D/\pa x')(x'_0,x_0'')=0$ and
$\op{det}(\pa^2D/\pa x' \pa x')(x_0', x_0'')\ne 0$. By the implicit function
theorem, Eq.~\thetag{1.12} has a unique smooth solution $x'=x'(x'')$ in a
neighborhood of $x_0$, and this solution is obviously the solution to the
minimization problem
\begin{equation}
D(x',x'')\to\min,\quad x''\text{ is fixed}, \,\, (x',x'') \text{ lies in a
neighborhood of } x_0.
\tag{1.13}
\end{equation}
Set
\begin{equation}
d(x'')\overset{\op{def}} = D(x'(x''),x'').
\tag{1.14}
\end{equation}
By construction, we have
\begin{equation}
d(x'')\leqslant D(x',x'')
\tag{1.15}
\end{equation}
in a neighborhood of $x_0$, and moreover,
\begin{equation}
D(x)\sim d(x'') + (x'-x'(x''))^2.
\tag{1.16}
\end{equation}
Let us now find the functions $g_j(x'')$. The functions must satisfy the
system
\begin{equation}
f(x)=C(x)(x'-g(x''))+\Phi(x),
\tag{1.17}
\end{equation}
where $C(x)$ is a smooth $k\times k$ matrix function,
$$
f(x)={}^t(f_1(x),\dots,f_k(x)),\quad g(x'')={}^t(g_1(x''),\dots,g_k(x'')),
$$
and $\Phi(x)\in\mathcal{D}$. We seek $g(x'')$ in the form
\begin{equation}
g(x'')=x'(x'')+h(x''),
\tag{1.18}
\end{equation}
where $h(x'')=O(d(x'')^{1/2})$ are new unknown functions. By Morse's lemma,
we have
\begin{align}
f(x) &= f(x'(x''),x'') + \frac{\pa f}{\pa x'}(x'(x''),x'')(x'-x'(x''))
\nonumber\\
&\quad+\langle x'-x'(x''), \Psi(x)(x'-x'(x''))\rangle,
\tag {1.19}
\end{align}
where $\Psi(x)$ is  smooth. Substituting Eqs.~\thetag{1.18}
and \thetag{1.19} into Eq.~\thetag{1.17} we obtain
\begin{align}
&f(x'(x''),x'') + \frac{\pa f}{\pa x'}(x'(x''),x'')(x'-x'(x''))
\nonumber\\
&\qquad\qquad
=C(x)(x'-x'(x''))-C(x)h(x'')+O( D(x)).
\tag{1.20}
\end{align}
We can satisfy Eq.~\thetag{1.20} by setting
\begin{equation}
C(x)=\frac{\pa f}{\pa x'}(x'(x''),x'')
\tag{1.21}
\end{equation}
(in particular, $C(x)$ is independent of $x'$) and
\begin{equation}
h(x'') = -\Big[\frac{\pa f}{\pa x'}(x'(x''),x'')\Big]^{-1}f(x'(x''),x'')
\tag{1.22}
\end{equation}
(note that $(\pa f/\pa x')(x)$ is invertible near $x_0$ by \thetag{1.8}).
Since $f(x'(x''),x'')=O(d(x'')^{1/2})$, the same is true of $h(x'')$. The
matrix $C(x)$ is invertible, and it follows from \thetag{1.17} that
$x'-g(x'')$ is a system of generators of $\mathcal{J}$. Finally,
$$
|x'-g(x'')|^2\leqslant |x'-x'(x'')|^2+|h(x'')|^2\leqslant CD(x);
$$
in conjunction with \thetag{1.16}, this implies that $D(x)\sim\wt D(x)$.

(c) Let $\wt g_1(x''),\dots,\wt g_k(x'')$ be  functions such that
$x_1'-\wt g_1(x''),\dots,x_k'-\wt g_k(x'')$ is a system of generators of
$\mathcal{J}$. Then there exists a nondegenerate matrix $\mathcal{E}(x)$ such that
\begin{equation}
x'-g(x'')=\mathcal{E}(x)(x'-\wt g(x''))+O(D).
\tag{1.23}
\end{equation}
Since
$$
\mathcal{E}(x)=\mathcal{E}(x'(x''),x'')+L(x)(x'-x'(x'')),
$$
we can assume that $\mathcal{E}(x)$ is independent of $x'$ in Eq.~\thetag{1.23},
i.e.,
\begin{equation}
x'-g(x'')=\mathcal{E}(x'')(x'-\wt g(x''))+O(D).
\tag{1.24}
\end{equation}
By differentiating \thetag{1.24} with respect to $x'$,
we obtain $\mathcal{E}(x'')=I+O(D^{1/2})$, where $I$ is the identity matrix, so that
$$
g(x'')-\wt g(x'') = O( D(x)),
$$
and, by minimizing with respect to $x'$, we obtain
\begin{equation}
g(x'')-\wt g(x'') = O(d(x'')).
\tag{1.25}
\end{equation}

Let us prove that $d(x'')$ is unique modulo equivalence; to this end, we take
some dissipation $D$ associated with $\mathcal{D}$ and note that the conditions
imposed in item (b) imply
\begin{align*}
&c\Big(d(x'') +\sum_{j=1}^k |x_j-\Re g_j(x'')|^2 + \sum_{j=1}^k|\Im
g_i(x'')|^2\Big)\leqslant D(x)
\\
&\leqslant C\Big(d(x'') +\sum_{j=1}^k|x_j-\Re g_j(x'')|^2 + \sum_{j=1}^k
|\Im g_j(x'')|^2 \Big)
\leqslant C_1\Big(d(x'') + \sum_{j=1}^k |x_j-\Re g_j(x'')|^2\Big),
\end{align*}
where $c, C, C_1$ are some positive constants, the last inequality
being due to the fact that $\Im g_j(x'')=O(d(x''))$. Let us pass to the
minimum over $x'$ in these inequalities and discard the nonnegative terms
under the summation signs on the left-hand side. We obtain
$$
cd(x'') \leqslant \min_{x'} D(x',x'') \leqslant C_1 d(x''),
$$
which implies that any function $d(x'')$ satisfying the conditions of item
(b) is equivalent to $\min D(x',x'')$.

(d) Let $\Phi(x)\in C^\infty(M)$. In the preceding notation we have
\begin{align*}
\Phi(x) &\equiv \Phi(x',x'') =\Phi(x'(x''),x'') +
(x'-x'(x''))\Phi_{x'}(x'(x''),x'')+ O(D)
\\
&=\Phi(x'(x''),x'') +[(g(x'')-x'(x''))+ (x'-g(x''))]\Phi_{x'}(x'(x''),x'')
+O(D);
\end{align*}
thus, the desired identity holds with
$$
\ph(x'') = \Phi(x'(x''),x'') + (g(x'')-x'(x''))\Phi_{x'}(x'(x''),x'');
$$
if some other function $\ph_1(x'')$ satisfies the same identity, then
$\ph(x'')-\ph_1(x'')\in{\mathcal{J}}$, and hence
\begin{align*}
\ph(x'') - \ph_1(x'') &= \sum_{j=1}^k a_j(x',x'')(x_j-g_j(x''))+O(D)
\\
&=\sum_{j=1}^k a_j(x'(x''),x'')(x_j-g_j(x''))+O(D).
\end{align*}
Differentiating both sides with respect to $x'$ yields
$a_j(x'(x''),x'')=O(D^{1/2})$; hence,
$
\ph(x'')-\ph_1(x'') = O(D),
$
and it remains to minimize the right-hand side over $x'$.

The theorem is thereby proved.
\qed \medskip

A trivial analog of Theorem 1.10 holds for complex coordinates.
\medskip

\noindent
{\bf Theorem 1.10$'$.}
{\it Let $L=(\mathcal{D}, \mathcal{J})$ be an asymptotic
submanifold of codimension $k$ in $M$, and let $(F_1,\dots,F_k,
F_{k+1},\dots,F_n) = (F',F'')$ be a complex coordinate system in a
neighborhood of a point $x_0\in\Ga$. Suppose that
$$
\op{det}\frac{\pa(f_1,\dots,f_k)}{\pa(F_1,\dots,F_k)}(x_0)\ne0.
$$
Then there exist functions $\Phi_1,\dots,\Phi_k$ such that
\begin{enumerate}
\item[{\rm(i)}] $F_1-\Phi_1,\dots, F_k-\Phi_k$ generate $\mathcal{J}$;
\item[{\rm(ii)}] $\pa\Phi_i/\pa F_j\in\mathcal{D}$, $i,j=1,\dots,k$.
\end{enumerate}
Condition {\rm(ii)} states that the functions $\Phi_i$ ``do not depend" on
$F_j$, $i,j=1,\dots,k$.}
\medskip

\noindent{\it Proof}.
Assuming summation over repeated indices from $1$ to $k$, we set
$$
\Phi_i=F_i-\Big(\frac{\pa f}{\pa F'}\Big)_{is}^{-1}f_s +
\frac12 f_l A_{ls}^i f_s,
$$
where
$$
A_{ls}^i = \Big(\frac{\pa f}{\pa F'}\Big)_{rl}^{-1}
\Big(\frac{\pa f}{\pa p'}\Big)_{im}^{-1}
\Big(\frac{\pa f}{\pa F'}\Big)_{ts}^{-1}
\frac{\pa^2 f_m}{\pa F_r \pa F_t};
$$
then
$$
\frac{\pa\Phi_i}{\pa F_s} = \frac12 \Big\langle
f,\frac{\pa A^i}{\pa F_s}f\Big\rangle\in\mathcal{D},\quad i,s=1,\dots,k,
$$
and
$$
F'-\Phi=
\Big(\frac{\pa f}{\pa F'}\Big)^{-1}f-\frac12\langle f, Af\rangle
$$
is obviously a system of generators of $\mathcal{J}$. The theorem is proved.
\qed \medskip

\subsection{Asymptotic submanifolds. Parametric local description}

There is still another way to describe asymptotic submanifolds, namely, by
using equations describing the ``embedding" of this manifold in $M$. Let
$U\subset \R^m$ be some domain, and let a dissipation $d(\a)$, $\a\in U$,
be given on $U$. Suppose that we have a set of functions
\begin{equation}
X(\a)=(X_1(\a),\dots, X_n(\a))
\tag{1.26}
\end{equation}
defined on $U$ such that the following conditions are satisfied:
\begin{align}
&i)\qquad\qquad\qquad\op{rank}_{\C} \Big(\frac{\pa X_1}{\pa\a},
\dots,\frac{\pa X_n}{\pa\a} \Big)=m,
\tag{1.27}\\
&ii)\qquad\qquad\qquad
\Im X_i(\a)=O(d(\a)^{1/2}).
\tag{1.28}
\end{align}
In particular, we have $m\leqslant n$, and only the case $m<n$ is of interest.

Let $(x_1,\dots,x_n)$ be a coordinate system about some point $m_0\in M$.
Then we can use the vector function \thetag{1.26} to define an asymptotic
submanifold in $M$ near $m_0$ as follows. Let $\a_0\in\Ga_d$, and let
$x_0=X(\a_0)$ (note that $x_0$ is necessary real). Set
\begin{equation}
D(x,\a)=d(\a)+\sum_{i=1}^n |x_i-X_i(\a)|^2
\tag{1.29}
\end{equation}
and consider the functions
\begin{equation}
\ph_i(x,\a) = x_i-X_i(\a),\quad i=1,\dots,n.
\tag{1.30}
\end{equation}
The pair $(\mathcal{D}, \mathcal{J})$, where $\mathcal{D}$ is the ideal corresponding to the
dissipation \thetag{1.29} and $\mathcal{J}$ is the ideal generated by $\mathcal{D}$ and
the functions \thetag{1.30}, is obviously an asymptotic submanifold in
$M\times U$ near $(x_0,\a_0)$. Let us construct the projection of this
submanifold on $M$ (this is possible by virtue of condition \thetag{1.27},
but the reader should be careful to keep in mind that the construction is
purely local).

We proceed as follows. By condition \thetag{1.27}, we have
\begin{equation}
\frac{\pa^2 D(x,\a)}{\pa\a \pa\a}\Bigg|_{x=x_0,\,\a=\a_0}>0
\tag{1.31}
\end{equation}
and hence the same is true in  a neighborhood of $(x_0,\a_0)$. Furthermore,
we have
\begin{equation}
\frac{\pa D}{\pa\a}(x_0,\a_0)=0,
\tag{1.32}
\end{equation}
and by the implicit function theorem there exists a smooth vector function
\begin{equation}
\a=\a(x),\quad \a(x_0)=\a_0,
\tag{1.33}
\end{equation}
that satisfies Eq.~\thetag{1.32} in a neighborhood of $x_0$. Set
\begin{equation}
\wt D(x)=D(x,\a(x)).
\tag{1.34}
\end{equation}
Obviously, $\wt D(x)\leqslant D(x,\a)$ in a neighborhood of $(x_0,\a_0)$.
Furthermore, set
\begin{equation}
\wt{\mathcal{J}} =\{f(x)\mid f(x)\otimes 1(\a)\in\mathcal{J}\}.
\tag{1.35}
\end{equation}
It is easy to see that the pair $(\wt{\mathcal{D}}, \wt{\mathcal{J}})$, where
$\wt{\mathcal{D}}$ is the ideal generated by $\wt D$ \thetag{1.34}, is an
asymptotic submanifold in $M$.

\subsection{Asymptotic mappings}

Let $L=(\mathcal{D},\mathcal{J})$ be a $k$-codimensional asymptotic submanifold in
$M$, and let $f:M\to N$ be a diffeomorphism. Then the image $\wt L=f(L)$ can be
defined in a natural manner as follows. We set $\wt{\mathcal{D}}=(f^{-1})^*\mathcal{D}$
and $\wt{\mathcal{J}}=(f^{-1})^*\mathcal{J}$, where $(f^{-1})^*$ acts elementwise,
that is, $(f^{-1})^*(\mathcal{A})=\{(f^{-1})^*\ph,\,\,\ph\in\mathcal{A}\}$. Then
$\wt L=(\wt {\mathcal{D}},\wt{\mathcal{J}})$. There is still another description of
$f(L)$.  In the Cartesian product $M\times N$ consider the submanifold
$$
\op{graph} f=\{(x,y)\in M\times N\mid y=f(x)\}.
$$
The associated asymptotic submanifold $({\mathcal{D}}_f,{\mathcal{J}}_f)$ in $M\times
N$ (cf. Example~1.8) can be described locally as follows. Let
$(x_1,\dots,x_n)$ and $(y_1,\dots,y_n)$ be local coordinate systems on $M$
and $N$, respectively, and let $f$ be given by the functions
\begin{align*}
y_1 &= f_1(x_1,\dots,x_n),\\
&\dots \quad\dots\quad\dots\\
y_n &= f_n(x_1,\dots,x_n)
\end{align*}
in the coordinates. Then ${\mathcal{D}}_f$ corresponds to the dissipation
$D_f(x,y)=\sum (y_j-f_j(x))^2$ and ${\mathcal{J}}_f$ is the ideal generated by
$y_1-f_1(x),\dots,y_n-f_n(x)$. It is an easy exercise to verify that if $D$
is a dissipation corresponding to $\mathcal{D}$, then $\wt{\mathcal{D}}$ is
associated with the dissipation
$$
\wt D(y) =\min_x \{D(x)+D_f(x,y)\}
$$
and that $\wt{\mathcal{J}}$ can be described as
$$
\wt{\mathcal{J}} =\{\ph(x,y)\in\mathcal{J} +{\mathcal{J}}_f\mid \ph\,\,\text{ is
independent of $x$}\}.
$$
Indeed, first of all, note that if $D_1(x,y)$ and $D_2(x,y)$ are
equivalent dissipations on $M\times N$, then the dissipations
$d_1(y)=\min_x D_1(x,y)$ and $d_1(y)=\min_x D_2(x,y)$ are also
equivalent; to observe this, it suffices to apply $\min_x$ to the
inequalities
$$
cD_1(x,y)\leqslant D_2(x,y)\leqslant CD_1(x,y).
$$
Thus, in the definition of $\wt D(y)$ we can safely replace
$D(x)+D_f(x,y)$ by
$$
D(x,y)=D(x)+\frac12 \sum(x_j-g_j(y))^2,
$$
where $x=g(y)$ is the inverse of the mapping $y=f(x)$.

Consider any point $x_0\in\Ga_{D(x)}$ and set $y_0=f(x_0)$. The mapping
$$
\si(x)=x+\frac{\pa D(x)}{\pa x}
$$
is an almost-identity diffeomorphism in a neighborhood of $x_0$. Indeed,
$$
\si(x)-x=\frac{\pa D(x)}{\pa x}=O(\sqrt{\mathcal{D}(x)}\,),
$$
and
$$
\frac{\pa\si}{\pa x}\Bigg|_{x=x_0}= I+
\frac{\pa^2D}{\pa x\pa x}\Bigg|_{x=x_0}>0,
$$
which implies that $\op{det}\pa\si/\pa x\ne0$ near $x_0$. For $y$ close
to $y_0$, the point
$$
y(x)=\op{arg}\min_x \Big\{D(x)+\frac12 \sum(x_j-q_j(y))^2\Big\}
$$
is determined from the equation
$$
\frac{\pa}{\pa x}\Big\{D(x)+\frac12\sum(x_j-g_j(y))^2\Big\}
= \si(x) - g(y)=0.
$$
That is, $x=\si^{-1}(g(y))$ and
$$
\wt D(y) = D(\si^{-1}(g(y)))+\frac12\|\si^{-1}(g(y))-g(y)\|^2.
$$
By using Lemma~1.6, we easily obtain $\wt D(y)\sim D(q(y))$.
Furthermore, $\ph(y)\in\wt{\mathcal{J}}$ if and only if $\ph(f(x))\in\mathcal{J}$.
We have
\begin{equation}
\ph(y)=\ph(f(x))+ C(x,y)(y-f(x))
\tag{1.36}
\end{equation}
by Hadamard's lemma (here $C(x,y)=(C_1(x,y),\dots,C_n(x,y))$ is a smooth
vector function). If $\ph(f(x))\in\mathcal{J}$, then it follows from
\thetag{1.36} that
\begin{equation}
\ph(y)\in\{\psi(x,y)\in{\mathcal{J}}+{\mathcal{J}}_f\mid\psi\,\,\text{ is independent
of $x$}\}.
\tag{1.37}
\end{equation}
Conversely, let \thetag{1.37} be true; then for some $a(x)\in\mathcal{J}$ we have
$$
\ph(y)=a(x)+b(x,y)(y-f(x)),
$$
and we find that $a(x)=\ph(f(x))$ by setting $y=f(x)$ in the last equation.

These considerations motivate the following definition.

\begin{DefinitioN} 
\rm
Let $M$ and $N$ be two manifolds of the same
dimension $n$, and let $G=({\mathcal{D}}_G, {\mathcal{J}}_G)$ be an $n$-dimensional
asymptotic submanifold in $M$ such that the following conditions are
satisfied:
\begin{enumerate}
\item[{(i)}] for any $x\in M$ there is at most one point $y\in N$ such that
$(x,y)\in\Ga_G\equiv\Ga_{{\mathcal{D}}_G}$, and {\it vice versa\/};
\item[{(ii)}] for any $(x_0,y_0)\in\Ga_G$ the Jacobians
$$
\op{det}\frac{\pa(f_1,\dots,f_n)}{\pa(x_1,\dots,x_n)}\quad\text{and}\quad
\op{det}\frac{\pa(f_1,\dots,f_n)}{\pa(y_1,\dots,y_n)}
$$
are nonzero for a certain (and hence for any) system $f_1,\dots,f_n$ of
generators of the ideal ${\mathcal{J}}_G$.
\end{enumerate}

Then $G$ is called a ({\it graph of\/}) {\it asymptotic diffeomorphism\/}
from $N$ into $M$ and is denoted $G:N\to M$.
\end{DefinitioN}

We are interested in the action of asymptotic diffeomorphisms on asymptotic
submanifolds.

\begin{Theorem} 
Let $L=(\mathcal{D}, \mathcal{J})$ be an asymptotic submanifold of $N$, and let
$G=(\mathcal{D}_G, \mathcal{J}_G)$ be an asymptotic diffeomorphism from $N$ into $M$.
Let $D$ and $D_G$ be dissipations associated with $\mathcal{D}$ and $\mathcal{D}_G$.
Suppose that the set
$$
\wt\Ga = \{x\in M\mid \exists y\in N\,:\,\,\,y\in\Ga_{\mathcal{D}} \,\,\text{ and
}\,\, (x,y)\in\Ga_{\mathcal{D}_G}\}
$$
is nonempty and define a dissipation $\wt D(x)$ on $M$ in the vicinity of
$\wt\Ga$ by the formula
$$
\wt D(x)=\min_y\{D(y)+D_G(x,y)\}
$$
{\rm(}the minimum is taken over some neighborhood of $\Ga_{\mathcal{D}}${\rm)}.
Let $\wt{\mathcal{D}}$ be the dissipation ideal associated with $\wt D(x)$, and
set $$
\wt{\mathcal{J}} =\{\ph(x,y)\in\mathcal{J} +{\mathcal{J}}_f\mid \ph\,\,\text{ is
independent of $y$}\}.
$$
Then $\wt L=(\wt{\mathcal{D}}, \wt{\mathcal{J}})$ is an asymptotic submanifold in $M$
and $\op{dim}\wt L=\op{dim}L$.
\end{Theorem}
\medskip

\noindent{\it Proof}.
Let $(x_0,y_0)\in\Ga_{\mathcal{D}_G}$ and $y_0\in\Ga_{\mathcal{D}}$. We have
$$
\frac{\pa^2}{\pa y\pa y}\{D(y)+D_G(x,y)\} =\frac{\pa^2D}{\pa y\pa y} +
\frac{\pa^2D_G}{\pa y\pa y}>0
$$
at $(x_0,y_0)$, and so
$$
y(x)=\op{arg}\min_y\{p(y)+D_G(x,y)\}
$$
is a smooth mapping in the vicinity of $x_0$ and $y(x_0)=y_0$.

By the implicit function theorem (Theorem~1.10), we can choose a set of
generators of ${\mathcal{J}}_G$ of the form
$$
f_j(x,y) = y_j - g_j(x),\quad j=1,\dots,n.
$$

For brevity, in what follows we write
$$
Q(x,y)=D(y)+D_G(x,y).
$$

Let $\psi_1(y),\dots,\psi_k(y)$ be a system of generators of the ideal ${\mathcal{J}}$. Set
\begin{align*}
&\Xi_j(x,y) =\psi_j(y) +\frac{\pa\psi(y)}{\pa y}(g(x)-y),
\\
&\wt\psi_j(x) = \psi_j(y(x)) +\frac{\pa\psi_j}{\pa y}(y(x))(g(x)-y(x)) =
\Xi_j(x,y(x)).
\end{align*}
Obviously, $\Xi_j(x,y)\in\mathcal{J}+{\mathcal{J}}_G$. Furthermore, we have
$$
\frac{\pa\Xi_j}{\pa y}(x,y) = \frac{\pa^2\psi_j(y)}{\pa y^2}(g(x)-y).
$$
Consequently,
$$
\wt\psi_j(x) -\Xi_j(x,y) = (y(x)-y) \frac{\pa^2\psi_j(y)}{\pa y^2}(g(x)-y) +
O(\|y(x)-y\|^2).
$$
Since $\pa^2Q/\pa y\pa y>0$, it follows that $\|y(x)-y\|^2=O(Q)$ and thus
$\wt\psi_j(x)\in{\mathcal{J}}+{\mathcal{J}}_G$; since $\wt\psi_j(x)$ is independent of
$x$, we see that $\wt\psi_j(x)\in\wt{\mathcal{J}}$.

Next, $\th(x)\in\wt{\mathcal{J}}$ if and only if
$$
\th(x) = \sum b_l(y)\psi_l(y) + \sum a_j(x,y)(y_j-g_j(x))+O(Q(x,y)),
$$
where $b_l(y)$ and $a_j(x,y)$ are smooth functions.

Set
$$
\wt\th(x) = \sum\Big[b_l(y(x)) +\frac{\pa b_l}{\pa
y}(y(x))(g(x)-y(x))\Big]\wt\psi_l(x).
$$
Then
\begin{align*}
&\th(x)-\wt\th(x) = \sum b_l(y)[\psi_l(y)-\wt\psi_l(x)] \\
&\quad+ \sum\Big[b_l(y)-b_l(y(x)) -\frac{\pa b_l}{\pa
y}(y(x))(g(x)-y(x))\Big] \wt\psi_l(x) + \sum a_j(x,y)(y_j-g_j(x))\\
&= \sum b_l(y)[\psi_l(y)-\Xi_l(x,y)] \\
&\quad+ \sum\Big[b_l(y)-b_l(y) -\frac{\pa b_l}{\pa
y}(y)(g(x)-y)\Big] \Xi_l(x,y) +  a(x,y)(y-g(x)) + O(Q(x,y))\\
&= \sum b_l(y)\frac{\pa\psi_l(y)}{\pa y}(y-g(x)) + \sum\frac{\pa b_l}{\pa y}
(y-g(x))\psi_l(y) + a(x,y)(y-g(x)) + O(Q(x,y))\\
&= c(x,y)(y-g(x)) +O(Q(x,y)) = c(x,y(x))(y-g(x)) +O(Q(x,y)).
\end{align*}
Since the left-hand side of the last equation is independent of $y$, it
follows by differentiation with respect to $y$ that $c(x,y(x))=O(Q^{1/2})$.
Therefore
$$
\Th(x) =\wt\Th(x)+ O(Q(x,y)) = \wt\Th(x) + O(\wt D(x))
$$
and we see that $\wt\psi_1(x),\dots,\wt\psi_l(x)$ is a system of generators
of the ideal $\wt{\mathcal{J}}$. Furthermore, we have
\begin{align*}
d\wt\psi_j(x) &= \Big[\frac{\pa\Xi_j}{\pa x} +
\frac{\pa\Xi_j}{\pa y}\frac{\pa y(x)}{\pa x}\Big]_{y=y(x)}\,dx\\
&= \Big[\frac{\pa\psi_j}{\pa y}(y(x))\frac{\pa g(x)}{\pa x} +
\frac{\pa^2\psi_j}{\pa y^2}(y(x))(g(x)-y(x))\Big]\,dx.
\end{align*}
At $(x_0,y_0)$ we have
$$
d\wt\psi_j = \frac{\pa\psi_j}{\pa y}(y_0)\frac{\pa g}{\pa x}(x_0)\,dx
$$
and, since $d\psi_j$ are linearly independent and $(\pa g/\pa x)(x_0)$ is a
nondegenerate matrix, it follows that  $d\wt\psi_j$ are linearly
independent. Finally, from the formula defining $\wt\psi_j(x)$ we obtain
$$
\wt\psi_j(x) = O(Q(x,y(x))^{1/2}) = O(\wt{\mathcal{D}}(x)^{1/2}),
$$
and $\wt{\mathcal{J}}\subset\wt{\mathcal{D}}^{1/2}$. The inclusion
$\wt{\mathcal{D}}\subset\wt{\mathcal{J}}$ is obvious.

The theorem is proved.
\qed \medskip

The asymptotic manifold $\wt L$ constructed in Theorem~1.12 will be denoted
$\wt L=G(L)$.

\begin{Theorem} 
Let $G:M_1\to M_2$ and $H:M_2\to M_3$ be asymptotic diffeomorphisms. Then
\begin{enumerate}
\item[{\rm(i)}] for any asymptotic submanifold $L$ in $M_1$ we have
$$
H(G(L)) = (H\circ G)(L),
$$
where $H\circ G: M_1\to M_3$ is the asymptotic diffeomorphism defined as
follows. Let $G=(\mathcal{D}_G,\mathcal{J}_G)$ and $H=(\mathcal{D}_H,\mathcal{J}_H)$, and let
$D_G(x,y)$ and $D_H(y,z)$ be dissipations associated with $\mathcal{D}_G$ and
$\mathcal{D}_H$, respectively. Then ${\mathcal{D}}_{H\circ G}$ is the dissipation
ideal corresponding to the dissipation
$$
{\mathcal{D}}_{H\circ G}(x,z) = \min_y \{D(x,y)+D(y,z)\},
$$
and
$$
{\mathcal{J}}_{H\circ G} = \{\ph(x,y,z)\in {\mathcal{J}}_{H} +
{\mathcal{J}}_{ G}\mid \ph\,\,\text{ is independent of } y\};
$$
\item[{\rm(ii)}] the composition of asymptotic diffeomorphisms thus defined
is associative, $(G\circ H)\circ K=G\circ(H\circ K)$.
\end{enumerate}
\end{Theorem}

We omit the proof of Theorem~1.13, since it is purely technical and contains
no new ideas as compared with the preceding theorem.

\section{Objects on asymptotic manifolds}

\subsection{Functions, vector fields, and differential forms}

Let $L=({\mathcal{D}}, {\mathcal{J}})$ be  a $k$-codimensional asymptotic submanifold
in $M$. We set
$$
\mathcal{C}^\infty(L)=\mathcal{C}^\infty(M)/\mathcal{J}
$$
and
$$
\mathcal{C}^\infty_{(1)}(L)=\mathcal{C}^\infty(M)/{\mathcal{D}}^{1/2}
= \mathcal{C}^\infty(L)/({\mathcal{D}}^{1/2}/\mathcal{J}).
$$
The sheaf $\mathcal{C}^\infty(L)$ is called the {\it sheaf of smooth functions\/} on
$L$. The reason for introducing the sheaf $\mathcal{C}^\infty_{(1)}(L)$ will be
clarified later on.
We shall also make extensive use of the sheaves
$$
\overset{\circ} {\mathcal{C}}^\infty(L) ={\mathcal{C}}^\infty(M)/
\overset{\circ} {\mathcal{J}},\qquad
\overset{\circ} {\mathcal{C}}_{(1)}^\infty(L)
={\mathcal{C}}^\infty(M)/\overset\circ {\mathcal{D}}^{1/2},
$$
where $\overset{\circ} {\mathcal{J}}=\mathcal{J}+\overset{\circ} {\mathcal{D}}$.

There is an obvious restriction map
$
i^*:\mathcal{C}^\infty(M)\to \mathcal{C}^\infty(L)
$
and also a projection
$
\pi: \mathcal{C}^\infty(L) \to \mathcal{C}^\infty_{(1)}(L),
$
such that $\mathcal{C}^\infty_{(1)}(L)$ possesses the natural structure of a
$\mathcal{C}^\infty(L)$-module. Similar mappings are defined for the ``circled"
spaces.

\begin{DefinitioN} 
\rm
A {\it vector field\/} on $L$ is a derivation
$X: \mathcal{C}^\infty(L)\to \mathcal{C}^\infty_{(1)}(L)$,
that is, a linear mapping such that
$$
X(fg) = X(f)\pi(g) + \pi(f)X(g)
$$
for any $f,g\in \mathcal{C}^\infty(L)$.
\end{DefinitioN}

\begin{LemmA} 
Let $Y:\mathcal{C}^\infty(M)\to \mathcal{C}^\infty(M)$ be a vector field on $M$, and
suppose that $Y\mathcal{J}\subset {\mathcal{D}}^{1/2}$. Then $Y$ correctly defines a
vector field on $L$.
\end{LemmA}

\noindent{\it Proof}.
By the hypotheses of the lemma, $Y$ factors through the natural projections
$$
\mathcal{C}^\infty(M)\to \mathcal{C}^\infty(M)/{\mathcal{J}},\qquad
\mathcal{C}^\infty(M)\to \mathcal{C}^\infty(M)/{\mathcal{D}}^{1/2}
$$
and thus gives rise to a vector
field on $L$.
\qed \medskip

\begin{DefinitioN} 
\rm
A vector field $Y$ on $L$ is said to be {\it geometric\/} if it is
obtained as a restriction of some vector field on $M$ (i.e., by the method
described in Lemma~2.2). The sheaf of geometric vector fields on $L$
will be denoted $\op{Vect}(L)$.
\end{DefinitioN}

Obviously, $\op{Vect}(L)$ is a ${\mathcal{C}}^\infty_{(1)}(L)$-module,
and we have
$$
\op{Vect}(L) = \op{Vect}_L(M)/{\mathcal{D}}^{1/2},
$$
where
$\op{Vect}_L(M)$ is the sheaf of vector fields satisfying the conditions
of Lemma~2.2 (such fields are said to be {\it tangent\/} to $L$). If
$X\in \op{Vect}_L(M)$ and $m_0\in\Ga$, then the vector $X(m_0)\in
T_{m_0}M$ is obviously well defined; such vectors will be called {\it
tangent vectors\/} to $L$ at $m_0$. The space of tangent vectors will be
denoted by $T_{m_0}M$. The following lemma shows that there are
``sufficiently many" geometric fields on $L$.

\begin{LemmA} 
Let $L=(\mathcal{D}, \mathcal{J})$ be a $k$-codimensional asymptotic submanifold in
an\linebreak $n$-dimensional manifold $M$. Then in a neighborhood of any
point $m_0\in\Ga$ there exist exactly $n-k$ geometric vector fields on $L$
linearly independent over $C^\infty_{(1)}(L)$.
\end{LemmA}

\noindent{\it Proof}.
Let us use the local description given in Theorem~1.10. Thus, we can assume
that the ideal $\mathcal{J}$ is generated by $\mathcal{D}$ and by the $k$ functions
$
x_i-g_i(x''),\quad i=1,\dots,k,
$
where $x=(x',x'')$, $x'=(x_1,\dots,x_k)$, $x''=(x_{k+1},\dots,x_n)$.

By the same theorem, an arbitrary function $f(x)=f(x',x'')\in
\mathcal{C}^\infty(M)$ can be represented in the form
\begin{equation}
f(x',x'')=f_0(x'')+\sum_{i=1}^k(x_i-g_i(x''))f_i(x'')+\eta(x),
\tag{2.1}
\end{equation}
where
\begin{align*}
f_0(x'') &= f(\Re  g'(x''),x'') + i\Bigg\langle \Im g'(x''), \frac{\pa f}{\pa
x'}(\Re g'(x''),x'')\Bigg\rangle,\\
f_i(x'') &= \frac{\pa f}{\pa x_i}(\Re g'(x''),x''),
\end{align*}
and $\eta(x)\in\mathcal{D}$.
Set
$$
X_j = \frac{\pa}{\pa x_j}-\sum_{i=1}^k \frac{\pa g_i(x'')}{\pa x_j}
\frac{\pa}{\pa x_i},\quad j=k+1,\dots,n.
$$
Then
$
X_j(x_i-g_i(x))=0,\quad i=1,\dots,k,
$
so that the operators $X_j$ give rise to geometric vector fields on $L$.
Next,
$$
X_j f(x',x'') = \frac{\pa f_0(x'')}{\pa x_j} +O(D^{1/2}),\quad
j=1,\dots,k.
$$
Let $a_j(x)$, $j=k+1,\dots,n$, be functions such that $\sum_{j=k+1}^n a_j(x)
X_j$ is the zero vector field on $L$. Then
$$
\sum_{j=k+1}^n a_j(x) \frac{\pa f_0}{\pa x_j}(x'') = O({ D}^{1/2})
$$
for any smooth function $f_0(x'')$. By choosing $f_0(x'')=x_j$, we see that
$a_j(x)\in{\mathcal{D}}^{1/2}$, $j=1,\dots,n$, that is, the $a_j(x)$ generate zero
elements in $\mathcal{C}^\infty_{(1)}(L)$.

Thus, the fields $X_j$ are linearly independent over  $\mathcal{C}^\infty_{(1)}(L)$.
Let us now prove that any system of $n-k+1$ vector fields is linearly
dependent over $\mathcal{C}^\infty_{(1)}(L)$.  This statement is obvious  from
the representation \thetag{2.1}. Indeed, let $X_1,\dots,X_{s+1}$ be
such a system (here $s=n-k$); then
$
X_i (x',x'') = X_i f_0(x'')\quad\text{in}\,\,\, \mathcal{C}^\infty_{(1)}(L)
$
and
$
X_i f_0(x'')=Y_i f_0(x''),
$
where $Y_i$ are some $s$-dimensional vector fields depending on
the parameters $x'$. However, the linear dependence of $Y_i$ over
$\mathcal{C}^\infty(M)$ is obvious, and hence the statement of the lemma
follows.
\qed \medskip

\noindent{\bf Remark.}
Lemma~2.4 can be restated as follows: $\op{Vect}(L)$ is a locally free
${\mathcal{C}}^\infty_{(1)}(L)$-module of rank $\op{dim}L$.
\medskip

\begin{DefinitioN} 
\rm
Let $X$ be a vector field tangent to $L$, $X\in\op{Vect}_L(M)$, and
suppose that the dissipation ideal $\mathcal{D}$ is invariant by $X$
and $X\overset{\circ}{\mathcal{J}}\subset
\overset{\circ}{\mathcal{J}}$. Then we say that $L$ is {\it strongly
invariant with respect to\/} $X$ (or $X$ is a {\it strong tangent field\/} to $L$).
\end{DefinitioN}

If $X$ is a strong tangent field to $L$, then $X$ acts as a derivation
of the sheaves
$$
X: \overset{\circ}{\mathcal{C}}^\infty(L)
\to \overset{\circ}{\mathcal{C}}^\infty(L)\quad\text{and}\quad
X:\overset{\circ}{\mathcal{C}}^\infty_{(1)}(L)\to
\overset{\circ}{\mathcal{C}}^\infty_{(1)}(L).
$$

\subsection{Differential forms}

\begin{DefinitioN} 
\rm
A {\it differential\/} $1$-{\it form\/} on $L$
is a ${\mathcal{C}}^\infty_{(1)}(L)$-linear functional $\om:\op{Vect}L\to
{\mathcal{C}}^\infty_{(1)}(L)$.
\end{DefinitioN}

The sheaf of differential $1$-forms on $L$ will be denoted by $\La^1(L)$;
by virtue of the preceding results, $\La^1(L)$ is a locally free
${\mathcal{C}}^\infty_{(1)}(L)$-module of rank $\op{dim}L$.

There is an obvious mapping $d:{\mathcal{C}}^\infty(L)\to\La^1(L)$; it is
given by the formula $df(X)\overset{\text{def}} = X(f)$; one can prove
that $\La^1(L)$ is generated over ${\mathcal{C}}^\infty(L)$ by elements of the
form $df$.

Furthermore, we have the commutative diagram
$$
\begin{CD}
{\mathcal{C}}^\infty(M) @>d>> \La^1(m)\\
@VVV  @VV i^* V\\
{\mathcal{C}}^\infty(\mathcal{L}) @>>d> \La^1(\mathcal{L})
\end{CD},
$$
where the left vertical arrow is the natural projection and
$$
i^*\om(X)=\om(\wt X)
$$
for any $\om\in\La^1(M)$ and any $x\in\op{Vect}(L)$, where $\wt
X\in\op{Vect}_L(M)$ is a representative of $X$. Since $X\in{\mathcal{D}}^{1/2}
\op{Vect}(M)$ implies $\om(X)\in{\mathcal{D}}^{1/2}$, it follows that $i^*$
is well defined.

\begin{DefinitioN} 
\rm
A {\it differential\/} $s$-{\it form\/} on $L$ is an alternating
$\mathcal{C}^\infty_{(1)}(L)$-polylinear mapping
$$
\om:\underbrace{\op{Vect}(L)\times\dots\times\op{Vect}(L)}_{s\text{ factors}}
\to \mathcal{C}^\infty_{(1)}(L).
$$
\end{DefinitioN}

We note that the mapping $i^*:\La^k(M)\to\La^k(L)$ is well defined for
any $k$.

We shall be mainly interested in $m$-forms, where $m=\op{dim}L$.
Nondegenerate $m$-forms will be referred to as volume forms. In this case,
the following assertion is valid.

\begin{LemmA} 
Let $\om$ be a differential $s$-form on an $s$-dimensional asymptotic
submanifold $L$ in $M$. Then in a neighborhood of each point $m_0\in\Ga$ the
form $\om$ is uniquely determined by its value on an arbitrary $s$-tuple
$(X_1,\dots,X_s)$ of linearly independent vector fields near $m_0$.
\end{LemmA}

The proof is obvious.

\begin{DefinitioN} 
\rm
Let $L=(\mathcal{D}, \mathcal{J})$ be an $s$-dimensional asymptotic submanifold in
$M$ and let $m_0\in\Ga$. A ({\it complex\/}) {\it coordinate system\/} on
$L$ in a neighborhood of $m_0$ is an $s$-tuple $(Q_1,\dots,Q_s)$ of
elements of ${\mathcal{C}}^\infty(L)$ such that $dQ_1\wedge\dots \wedge
dQ_s\ne 0$ at $m_0$ (or, which is the same, the differentials $dQ_1,\dots,
dQ_s$ are linearly independent near $m_0$).
\end{DefinitioN}

Sometimes we shall consider representatives $\wt Q_1,\dots,\wt Q_s$ of
$Q_1,\dots,Q_s$ in ${\mathcal{C}}^\infty(M)$; these will also be referred to
as local coordinates on $M$.

Since $dQ_1,\dots,dQ_s$ are linearly independent, we have a unique
decomposition
$$
df=a_1\,dQ_1 +\cdots+a_s\,dQ_s
$$
for any $f\in {\mathcal{C}}^\infty(L)$. The coefficients
$a_j\in {\mathcal{C}}^\infty_{(1)}(L)$ are denoted $a_j=\pa f/\pa Q_j$
and are referred to as
the partial derivatives of $f$ with respect to $Q_j$.

\begin{PropositioN} 
Let $L=(\mathcal{D}, \mathcal{J})$ be a $k$-codimensional submanifold in $M$, and
let $(Q_{k+1},\dots,Q_n)$ be a local coordinate system on $L$. Then
\begin{enumerate}
\item[{\rm(a)}] $\pa/\pa Q_{k+1},\dots, \pa/\pa Q_n\in\op{Vect}(L)$;
\item[{\rm(b)}] if $F_{k+1},\dots, F_n$ are arbitrary representatives of
$Q_{k+1},\dots, Q_n$ in ${\mathcal{C}}^\infty(L)$, then we can complete
$(F_{k+1},\dots,F_n)$ to a coordinate system on $M$ such that for any
$\ph\in {\mathcal{C}}^\infty(L)$ the following conditions are satisfied:

{\rm(b1)} $\pa\Phi/\pa F_j =\pa\ph/\pa Q_j$, $j=k+1,\dots,n,$ in
${\mathcal{C}}^\infty_{(1)}(L)$ for any representative $\Phi\in
{\mathcal{C}}^\infty(M)$ of $\ph$;

{\rm(b2)} there exists a representative $\wt\Phi\in
{\mathcal{C}}^\infty(M)$ of $\ph$ such that $\pa\Phi/\pa F_j\in\mathcal{D}$,
$j=1,\dots,k$.
\end{enumerate}
\end{PropositioN}
\medskip

\noindent{\it Proof}.
Let $(F_1,\dots,F_k)$ be a system of generators of $\mathcal{J}$, and let
$m\in\Ga$. Then $dF_i(\xi)=0$, $i=1,\dots,k$, for any $\xi\in T_m\Ga$.
Since $dF_1\wedge\cdots\wedge dF_k\ne0$ and since
$dF_{k+1}\wedge\cdots\wedge dF_n\ne0$ at $m$, $(F_1,\dots,F_n)$
is a coordinate system on $M$ in a neighborhood of $m$. Let $\Phi\in
{\mathcal{C}}^\infty(M)$ be an arbitrary representative of $\ph\in {\mathcal{C}}^\infty(L)$; then
$$
d\Phi=a_1\,dF_1+\cdots+ a_k\,dF_k +a_{k+1}\,dF_{k+1}+\cdots+
a_n\,dF_n.
$$
Since $F_1,\dots,F_k\in{\mathcal{J}}$, the first $k$ terms lie in the kernel
of $i^*:\La^1(M)\to\La^1(L)$, and (b1) is proved. Furthermore, we have
$\pa F_j/\pa F_s=0$ for $j\leqslant k<s$. Since $F_1,\dots,F_k$ span
$\mathcal{J}$ modulo $\mathcal{D}$, it follows that $\pa/\pa F_s(\mathcal{J})\subset
{\mathcal{D}}^{1/2}$, $s=k+1,\dots,n$, that is, the field $\pa/\pa F_s$ is
tangent to $L$. We see that $\pa/\pa Q_s$ are geometric vector fields on
$L$, generated by $\pa/\pa F_s$, and (a) is proved. Finally, we set
$$
\wt\Phi = \Phi-\sum_{j=1}^k F_j \frac{\pa\Phi}{\pa F_j} +\frac12
\sum_{j,s=1}^k F_j F_s \frac{\pa^2\Phi}{\pa F_j \pa F_s}.
$$
Then $\wt\Phi-\Phi\in\mathcal{J}$ and $\pa\wt\Phi/\pa F_j\in\mathcal{D}$, which
implies (b2). The proposition is proved.
\qed \medskip

\begin{DefinitioN} 
\rm
Let $\om$ be a volume form on an $s$-dimensional asymptotic submanifold
$L$, and let $Q_1,\dots,Q_s$ be coordinates on $L$. The function
$$
\frac{D\om}{DQ}\overset{\text{def}} = \om\Big(\frac{\pa}{\pa Q_1},
\dots,\frac{\pa}{\pa Q_s}\Big)\in \mathcal{C}^\infty_{(1)}(L)
$$
is called the {\it density\/} of $\om$ in the coordinates
$Q_1,\dots,Q_s$.
\end{DefinitioN}

Let $X\in\op{Vect}M$ be a strong tangent field to a $k$-codimensional
asymptotic submanifold $L$. Let $(Q_{k+1},\dots, Q_n)$  be an arbitrary
coordinate system on $L$, and let $(F_1,\dots,F_n)$ be any coordinate
system on $M$ constructed in the proof of Proposition~2.10 (that is,
$(F_1,\dots,F_k)$ is a $k$-tuple of generators of $\mathcal{J}$ and
$F_{k+1},\dots, F_n$ are representatives of $(Q_{k+1},\dots,Q_n)$). Then
$
X=\sum_{j=1}^n a_j({\pa}/{\pa F_j}),
$
and the strong tangency condition in particular implies that
$a_j\in\overset{\circ}{\mathcal{J}}$, $j=1,\dots,k$. Set
\begin{equation}
\op{div}_Q X =\sum_{j=k+1}^n \frac{\pa a_j}{\pa F_j}.
\tag{2.2}
\end{equation}

\begin{PropositioN} 
{\rm(a)} $\op{div}_Q X$ is a well-defined element of
$\overset{\circ} {\mathcal{C}}^\infty_{(1)}(L)$.

{\rm(b)} If $\wt Q=(\wt Q_{k+1},\dots, \wt Q_n)$ is another coordinate
system on $L$, then
\begin{equation}
\op{div}_{\wt Q} X=\op{div}_Q X+X\Big(\op{ln}\op{det}\frac{\pa\wt
Q}{\pa Q}\Big).
\tag{2.3}
\end{equation}
\end{PropositioN}
\medskip

\noindent{\it Proof}.
First, let us establish that the class of \thetag{2.2} in
$\overset{\circ} {\mathcal{C}}^\infty_{(1)}(L)$ does not depend on the choice of
$F_1,\dots,F_k$; to this end, let $S_i=\sum_{j=1}^k A_{ij}F_j +O(\mathcal{D})$,
$i=1,\dots,k$, be another set of generators of $\mathcal{J}$, and set
$S_i=F_i$, $i=k+1,\dots,n$. We have
\begin{equation}
\frac{\pa}{\pa F_j}=\sum_{l=1}^k
\frac{\pa S_l}{\pa F_j}\frac{\pa}{\pa S_l} =
\begin{cases}
\sum_{l=1}^k A_{lj}\frac{\pa}{\pa S_l} +O({\mathcal{D}}^{1/2}),\quad
1\leqslant j\leqslant k,\\
\frac{\pa}{\pa S_j} + O({\mathcal{D}}^{1/2}),\quad k+1\leqslant j\leqslant n.
\end{cases}
\tag{2.4}
\end{equation}
Consequently,
$$
X=\sum_{j,l=1}^k A_{lj} a_j \frac{\pa}{\pa S_l} +
\sum_{l=k+1}^n  a_l \frac{\pa}{\pa S_l} +O({\mathcal{D}}^{1/2}).
$$
Now we have
$$
\sum_{j=k+1}^n  \frac{\pa a_j}{\pa F_j}=
\sum_{j=k+1}^n   \frac{\pa a_j}{\pa S_j},
$$
as desired.

Let us now fix $F_1,\dots,F_k$ and consider some representatives
$(S_{k+1},\dots, S_n)$ of the coordinate system $\wt Q_{k+1},\dots,\wt
Q_n$. This time, we set $S_i=F_i$, $i=1,\dots,k$. We now have
\begin{equation}
\frac{\pa}{\pa F_j}=\sum_{l=1}^n
\frac{\pa S_l}{\pa F_j}\frac{\pa}{\pa S_l} =
\begin{cases}
\frac{\pa}{\pa S_j} +\sum_{k=l+1}^n
\frac{\pa S_l}{\pa F_j}\frac{\pa}{\pa S_l} ,\quad
1\leqslant j\leqslant k,\\
\sum_{k=l+1}^n
\frac{\pa S_l}{\pa F_j}\frac{\pa}{\pa S_l},\quad k+1\leqslant j\leqslant n,
\end{cases}
\tag{2.5}
\end{equation}
and so
$$
X=\sum_{l=1}^k a_l\frac{\pa}{\pa S_l} + \sum_{l=k+1}^n\sum_{j=1}^n
\frac{\pa S_l}{\pa F_j} a_j\frac{\pa}{\pa S_l}.
$$
We have
$$
\op{div}_{\wt Q}X =\op{div}_F X-\sum_{j=1}^k \frac{\pa a_j}{\pa F_j},
\quad\text{where }\quad
\op{div}_F X =\sum_{j=1}^n \frac{\pa a_j}{\pa F_j}.
$$
Similarly,
$$
\op{div}_{\wt Q}X =\op{div}_S X-\sum_{j=1}^k \frac{\pa a_j}{\pa S_j} =
\op{div}_S X- \sum_{j=1}^k  \frac{\pa a_j}{\pa F_j}
+O(\overset{\circ}{\mathcal{D}}^{1/2})
$$
by virtue of \thetag{2.5}, since $a_j\in\overset{\circ}{\mathcal{J}}$ for
$j=1,\dots,k$ and $\pa/\pa S_l$ is tangent to $L$ for $l=k+1,\dots,n$.
Thus,
$$
\op{div}_{\wt Q}X -\op{div}_Q X=\op{div}_S X-\op{div}_F X +
O(\overset{\circ}{\mathcal{D}}^{1/2})
= X\Big(\op{ln}\op{det}\frac{\pa S}{\pa F}\Big)
+O(\overset{\circ}{\mathcal{D}}^{1/2})
$$
(the last equality is valid by Sobolev's lemma; e.g., see \cite{MN}).
Since $S_i=F_i$ for $i=1,\dots,k$, it follows that
$$
\op{det}\frac{\pa S}{\pa F} =\op{det}\frac{\pa (S_{k+1},\dots,S_n)}{\pa
(F_{k+1},\dots,F_n)}.
$$
Thus, the class of $\op{det}\pa S/\pa F$ in $\overset{\circ}{\mathcal{C}}^\infty_{(1)}(L)$
is $\op{det}\pa\wt Q/\pa Q$, and since $X$ is a
strong tangent field, it follows that the class of $X(\op{ln}\op{det}\pa
S/\pa F)$ in $\overset{\circ}{\mathcal{C}}^\infty_{(1)}(L)$ is well defined and is equal to
$X(\op{ln}\op{det}\pa \wt Q/\pa Q)$.
We have thus arrived at \thetag{2.3}; by taking $\wt Q=Q$
in \thetag{2.3}, we see that $\op{div}_{\wt Q}X=\op{div}_{Q}X$, i.e., the
definition of $\op{div}_{ Q}X$ is independent of the choice of
$F_{k+1},\dots, F_n$. Proposition~2.12 is proved.
\qed \medskip

Now let $\op{dim}L=s$, and let $\om\in\La^s(L)$ be a volume form on $L$. In
an arbitrary system of local coordinates $Q_1,\dots,Q_s$ on $L$ we have
$$
\om=\frac{D\om}{DQ}dQ_1\wedge\cdots\wedge dQ_s.
$$

\begin{DefinitioN} 
\rm
Let $X$ be a strong tangent field to $L$. We define the {\it Lie
derivative of $\om$ along\/} $X$ by setting
\begin{equation}
\mathcal{L}_X\om
=
\Big[X\Big(\frac{D\om}{DQ}\Big) +
\frac{D\om}{DQ}\op{div}_QX\Big]\,dQ_1\wedge\cdots\wedge dQ_s.
\tag{2.6}
\end{equation}
\end{DefinitioN}

\begin{LemmA} 
Equation \thetag{2.6} specifies a well-defined element $\mathcal{L}_X\om\in
\overset{\circ}{\La}^s(L)$.
\end{LemmA}
\medskip

\noindent{\it Proof}.
We need to show that the form \thetag{2.6} is independent of the choice
of the coordinates $(Q_1,\dots,Q_s)$. If $(\wt Q_1,\dots,\wt Q_s)$ is
another system of coordinates on $L$, then we have ($D\wt
Q/DQ=\op{det}\pa\wt Q/\pa Q$)
\begin{align*}
&X\Big(\frac{D\om}{D\wt Q}\Big) + \frac{D\om}{D\wt Q} \op{div}_{\wt Q}X =
X\Big(\frac{D\om}{DQ}\frac{DQ}{D\wt Q}\Big) + \frac{D\om}{D\wt
Q}\Big(\op{div}_{Q}X + \Big(\frac{D\wt Q}{D
Q}\Big)^{-1}X\Big(\frac{D\wt Q}{D Q}\Big)\Big)
\\
&= \Big(\frac{D\wt Q}{D Q}\Big)^{-1}X
\Big(\frac{D\om}{D Q}\Big) +\frac{D\om}{D
Q}X\Big(\frac{D Q}{D\wt Q}\Big)  +
\frac{D \om}{D Q}\frac{D Q}{D\wt Q}
\Big(\op{div}_{Q}X + \Big(\frac{D\wt Q}{D
Q}\Big)^{-1}X\Big(\frac{D\wt Q}{D Q}\Big)\Big)
\\
&=\frac{D Q}{D\wt Q}\Big(X\Big(\frac{D\om}{D Q}\Big) +
\frac{D\om}{D Q}\op{div}_{Q}X
\Big) +\frac{D Q}{D\wt Q}\frac{D \om}{D Q}
\Bigg[\Big(\frac{D Q}{D\wt Q}\Big)^{-1}X
\Big(\frac{D Q}{D\wt Q}\Big) +
\Big(\frac{D \wt Q}{D Q}\Big)^{-1} X
\Big(\frac{D\wt Q}{D Q}\Big)
\Bigg].
\end{align*}
However, the terms in the square brackets cancel out, and we obtain the
desired result. Lemma~2.14 is proved.
\qed \medskip

\begin{DefinitioN} 
\rm
A volume form $\om\in\La^s(L)$ is said to be {\it invariant\/} with
respect to a strong tangent vector field $X$ if $\mathcal{L}_X\om=0$.
\end{DefinitioN}

In the sequel we also need the following technical result.

\begin{LemmA} 
Let $m_0\in\Ga$, $f\in C^\infty_{(1)}(L)$, $f(m_0)\ne0$ {\rm(}note that the
value of $f$ at $m_0$ is well defined\/{\rm)}. Then the square root $\sqrt
f$ is a well-defined element of $C^\infty_{(1)}(L)$ in a neighborhood of $m_0$.
\end{LemmA}

\noindent{\it Proof}.
Let $f_1,f_2\in C^\infty(M)$ be two representatives of $f$. Then
$f_1(m_0)=f_2(m_0)\ne 0$, $f_1-f_2\in{\mathcal{D}}^{1/2}$. We have
$$
\sqrt{f_2} = \sqrt{f_1+f_2-f_1} =
\sqrt{f_1}\sqrt{1+f_2-f_1}=\sqrt{f_1}+O({\mathcal{D}}^{1/2}).
$$
The lemma is proved.
\qed \medskip

\subsection{Bundles and connections}

\begin{DefinitioN} 
\rm
Let $L$ be an asymptotic submanifold in $M$, and let $E$ be a linear space
over $\C$. A  {\it vector bundle with fiber\/} $E$ {\it over\/} $L$ is a
$\mathcal{C}^\infty(L)$-module $\mathcal{E}$ on $M$ locally isomorphic to
$\mathcal{C}^\infty(L)\underset \C\otimes E$.
\end{DefinitioN}
\medskip

\noindent{\bf Remark.} What we have defined is in fact an analog of the sheaf of
germs of sections of a vector bundle.
\medskip

\noindent{\bf Example.}
The ``tangent bundle" $\op{Vect}(L)$ and the ``cotangent bundle"
$\La^1(L)$ are $s$-dimen\-sional vector bundles over $L$ (here
$s=\op{dim}L$).
\medskip

If $\mathcal{E}$ is a vector bundle with fiber $E$ over $L$, then we introduce
the sheaves
$$
{\mathcal{E}}_{(1)} = {\mathcal{E}}/{{\mathcal{D}}^{1/2}\mathcal{E}},\quad
\overset{\circ}{\mathcal{E}} = {\mathcal{E}}/\overset{\circ}{\mathcal{D}}\mathcal{E},\quad
\overset{\circ}{\mathcal{E}}_{(1)} = {\mathcal{E}}/
\overset{\circ}{\mathcal{D}}^{1/2}\mathcal{E}
$$
(note that the action of ${\mathcal{D}}^{1/2}$, $\overset{\circ}{\mathcal{D}}$, and
$\overset{\circ}{\mathcal{D}}^{1/2}$ on $\mathcal{E}$ is naturally defined). These
sheaves are $\mathcal{C}_{(1)}^\infty(L)$,  $\overset{\circ}{\mathcal{C}}^\infty(L)$,
and $\overset{\circ} {\mathcal{C}}_{(1)}^\infty(L)$-modules,
respectively, and there are natural homomorphisms
$$
\mathcal{E}\to \overset{\circ}{\mathcal{E}} \to{\mathcal{E}}_{(1)} \to
\overset{\circ}{\mathcal{E}}_{(1)}
$$
over the homomorphisms of sheaves of rings
$$
\mathcal{C}^\infty(L)\to
\overset{\circ}{\mathcal{C}}^\infty(L)\to
{\mathcal{C}}_{(1)}^\infty(L)\to
\overset{\circ}{\mathcal{C}}_{(1)}^\infty(L).
$$
Let $\pi:F\to M$ be a vector bundle with fiber $E$ over $M$.
Consider the sheaf $\mathcal{F}$ of germs of sections of $F$. If $L=(\mathcal{D},
\mathcal{J})$ is an asymptotic submanifold in $M$, then we can define the
{\it pullback\/} of $\mathcal{F}$ on $L$ by setting
\begin{equation}
\mathcal{E}=i^*\mathcal{F} = \mathcal{F}/\mathcal{J}\mathcal{F}
\tag{2.7}
\end{equation}
(note that $i^*$ in \thetag{2.7} symbolizes the pullback by the
``embedding" $i=L\hookrightarrow M$).

Any vector bundle $F$ over $M$ is a subbundle
of some trivial bundle $M\times B$, where $B$ is a
vector space over $\C$, and hence can be specified by a smooth
projection-valued mapping
\begin{equation}
\Pi: M\to \op{End}(B),
\tag{2.8}
\end{equation}
the range of $\Pi(x)$ being the fiber of $F$ over $x\i M$ (if the space $B$
is infinite-dimensional, the case which is important in applications, then
one should be very careful about the differentiability conditions to be
imposed on $\Pi$; in any case we assume that the range of $\Pi$ is
finite-dimensional).

The smooth sections of $F$ are the mappings $u:M\to B$ such that $\Pi(x)
u(x)=u(x)$ for any $x\in M$. Let $\mathcal{E}$ be the pullback \thetag{2.7}.
We shall briefly discuss the nonparametric local description of $\mathcal{E}$.

Let $x=(x';x'') =(x_1,\dots,x_k; x_{k+1},\dots,x_n)$ be a local
coordinate system on $M$, let $\mathcal{D}$ be the dissipation ideal
associated with the dissipation
$$
D(x)=d(x'') +\|x'-g(x'')\|^2,
$$
and let $\mathcal{J}$ be the ideal generated by $\mathcal{D}$ and by the functions
$x_i-g_i(x'')$, $i=1,\dots,k$\linebreak (cf. Theorem~1.10). Set
$$
x'(x'') =\op{arg}\min_{x'}D(x).
$$
By following the proof of Theorem~1.10, it is easy to establish that
$\Pi(x)$ and $u(x)$ can be represented in the form
$$
\Pi(x) = \wt\Pi(x'') +\Pi_1(x),\quad u(x)=\wt u(x'')+u_1(x),
$$
where $\Pi_1(x)\in\mathcal{J}\op{End}(B,B)$,
$u_1(x)\in\mathcal{J}{\mathcal{C}}^\infty(M,B)$, and $\wt\Pi$ and $\wt u$
are unique modulo $O(d(x''))$.
The explicit form of $\wt\Pi$ and $\wt u$ is given by
\begin{align}
\wt\Pi(x'') &= \Pi(x'(x''),x'' + \big[g(x'')-x'(x'')\big]\frac{\pa\Pi}{\pa x'} (x'(x''),x'');
\nonumber\\
\wt u(x'') &= u(x'(x''),x'' + \big[g(x'')-x'(x'')\big]\frac{\pa\wt u}{\pa
x'} (x'(x''),x'').
\tag{2.9}
\end{align}
The objects \thetag{2.9} will be referred to as the {\it local
representatives of \/} $\Pi$ {\it and\/} $u$ in the coordinates $x''$. A
straightforward calculation yields
$$
\wt\Pi^2=\wt\Pi+O(d),\qquad \wt\Pi\wt u = \wt u +O(d).
$$
If we regard the local representatives as classes modulo $O(d)$ rather
than functions, then we have $\wt\Pi^2=\wt\Pi$ and $\wt\Pi\wt u=\wt u$.

In the following we shall make some use of connections and covariant
derivatives.

\begin{DefinitioN} 
\rm
Let $\mathcal{E}$ be a vector bundle over $L$. A {\it connection\/} $\pa$ on
$\mathcal{E}$ is a $\C$-linear mapping
$$
\pa:\mathcal{E}\to\mathcal{E}_{(1)}\otimes\La^1(L)
$$
such that for any $f\in C^\infty(L)$ and any $\ph\in\mathcal{E}$ we have
\begin{equation}
\pa(f\ph) = f\pa\ph + \ph\otimes df
\tag{2.10}
\end{equation}
(we write $f$ instead of $\pi(f)$ on the right-hand side in \thetag{2.10}).
\end{DefinitioN}

Let $X$ be a vector field on $L$. Then the {\it covariant derivative\/}
$\nabla_X\ph$ of a section $\ph\in\mathcal{E}$ is defined as follows:
$$
\nabla_X \ph \overset{\text{def}} = \pa\ph(X).
$$
This is well defined, since $\pa\ph\in\mathcal{E}_{(1)}\otimes\La^1(L)$ and can
be applied to $X$ with respect to the second factor of the tensor product.

Let $F\subset M\times B$ be a subbundle of the trivial bundle $M\times
B$, and let $\Pi:M\to\op{End}B$ be the corresponding projection family.

The bundle $F$ is equipped with the natural Levi-Civit\`a connection
$
\pa =\Pi d.
$
It is easy to see that this connection factors
through the natural projections, so that we obtain a connection
$\wt\pa=i^*\pa$ on the pullback $\mathcal{E}=i^*\mathcal{F}$ \thetag{2.7}.
Obviously, in the local coordinates $x''$ we have $\pa=\wt\Pi\wt d$,
where $\wt d$ is the differential with respect to the local coordinates.

\section{Positive asymptotic Lagrangian submanifolds}

We begin by recalling, without proof, the notion and the main points
concerning Lagrangian asymptotic manifolds as defined in \cite{VDM1}.
Then we devise a new definition in the spirit of the approach outlined in
\S1 and \S2 and show that the two approaches are equivalent. This will
help us save space by resorting to some proofs that have already been
published.

But first of all, let us introduce some notation.

In this section we deal with asymptotic submanifolds in $\R^{2n}$. The
coordinates in $\R^{2n}$ will be denoted by
$(p,q)=(p_1,\dots,p_n,q_1,\dots,q_n)$. We assume that $R^n$ is equipped with
the standard symplectic structure
\begin{equation}
\om^2=dp\wedge dq \equiv \sum_{i=1}^n dp_i\wedge dq_i.
\tag{3.1}
\end{equation}
The following notation, in fact standard in the literature on the canonical
operator, will be used freely. Let $I=\{i_1,\dots,i_k\}\subset
\{1,\dots,n\}$ be an arbitrary subset. Then by $\ov I$ we denote its
complement $\ov I=\{1,\dots.n\}\setminus I= \{i_{k+1},\dots,i_n\}$,  by
$|I|$ the cardinality $|I|=k$, and if $\xi=(\xi_1,\dots,\xi_n)$ is an
$n$-vector, then $\xi_I$ is used to denote the $k$-vector
$(\xi_{i_1},\dots,\xi_{i_k})$ and $\xi_{\ov I} =
(\xi_{i_{k+1}},\dots,\xi_{i_n})$. Furthermore, we feel free to write
$p_I\,dq_I$ for $\sum_{i\in I} p_i\,dq_i$ etc; however, unless otherwise
specified, summation is {\it never\/} assumed in matrices of second partial
derivatives; thus, $\pa^2\Phi/\pa q_I \pa q_I$ may stand for the matrix
$(\pa^2\Phi/\pa q_i \pa q_j)_{i,j\in I}$ rather than for its trace; we even
sometimes write $\xi_I(\pa^2\Phi/\pa x_I \pa x_I)\xi_I$ to denote
$$
\sum_{i,j\in I}\xi_i\xi_j \frac{\pa^2\Phi}{\pa x_i \pa x_j},
$$
but if misunderstanding is likely to occur, then the less ambiguous
notation
$$
\Big\langle
\xi_I,\frac{\pa^2\Phi}{\pa x_I \pa x_I}\eta_I
\Big\rangle
$$
is used; here $\langle\cdot\,,\cdot\rangle$ is the standard bilinear pairing
of vectors in $\C^{|I|}$.

Finally, $dp_I\wedge dq_{\ov I}$ stands for
$(-1)^\si dp_{i_1}\wedge\dots\wedge dp_{i_k}\wedge dq_{i_{k+1}}\wedge\dots
\wedge dq_{i_n},
$
where $\si$ is the parity of the permutation $(i_1,\dots,i_n)$; the effect
is as if the factors were arranged in the ascending order of the
subscripts. The subscript $I$ is usually omitted altogether if
$I=\{1,\dots,n\}$.

For any $I\subset\{1,\dots,n\}$ we define a transformation
$\ga_I:\R^{2n}\to\R^{2n}$ by setting
\begin{equation}
\ga_I(p,q) = \big((p_I,-q_{\ov I}), (q_I,p_{\ov I})\big).
\tag{3.2}
\end{equation}
Note that
$\ga_I^*\om^2=\om^2$ and $\ga^*(p\,dq)=p_I\,dq_I-q_{\ov I}\,
dp_{\ov I}$.

\subsection{Lagrangian asymptotic manifolds:\\ one of the traditional definitions}

In this subsection we follow \cite{VDM1} and \cite{VDM2} with minor
alterations as to notation and the form of presentation. However, there
is one significant difference: here, as well as in \S1 and \S2, we deal
only with asymptotic Lagrangian submanifolds in the first approximation
($c$-Lagrangian structures in the terminology of \cite{VDM2}); the
accuracy $O(\mathcal{D}^\infty)$ is actually  redundant and is not used. We omit
all proofs, which can be extracted from \cite{VDM1} and \cite{VDM2}.

\begin{DefinitioN} 
\rm
A {\it Lagrangian chart\/} is a quintuple $r=(U,d,P,Q,W)$, where
$U\subset\R_\a^n$ is a domain and $d: U\to\R_+$; $P,Q:U\to\C^n$, $W:
U\to\C$ are smooth functions such that

i) $\op{rank}_{\C}\Big(\dfrac{\pa P(\a)}{\pa\a},
\dfrac{\pa Q(\a)}{\pa\a}\Big)=n$ for $\a\in \Ga_d$;

ii) $\Im P=O(d^{1/2})$, $\Im Q=O(d^{1/2})$, $\Im W=O(d)$;

iii) $(P,Q):\Ga_d\to\R^n$ is a topological embedding (note that
$(P,Q)|_{\Ga_d}$ are real by ii));

iv) $dW=P\,dQ+O(d)$.
\end{DefinitioN}

We denote $\Ga(r)=\{(p,q)\in\R^{2n}\mid p=P(\a),\,\, q=Q(\a)$ for some
$\a\in\Ga_\a$\}; the set $\Ga(r)$ is called the {\it zero image\/} of the
chart $r$. The function $W(\a)$ is called the {\it action\/} in the chart
$r$.

\begin{DefinitioN} 
\rm
Two  Lagrangian charts
$$
r=(U,d,P,Q,W)\quad\text{ and}\quad \wt r=(\wt U,\wt d,
\wt P,\wt Q,\wt W)
$$
are said to be {\it consistent\/} if for any two points $\a_0\in\Ga_d$
and $\wt\a_0\in\Ga_{\tilde d}\,$ such that $(P(\a_0), Q(\a_0))
=(\wt P(\wt\a_0), \wt Q(\wt\a_0))$ there exists a neighborhood
$V\subset U$ of $\a_0$, a neighborhood $\wt V\subset \wt U$ of $\wt\a_0$,
and a
diffeomorphism $V\leftrightarrow \wt V$, $\a_0\mapsto \wt\a_0$, such
that (under the identification of $\a$ with $\wt\a$ by this
diffeomorphism)

i) $d$ and $\wt d$ define the same dissipation ideal $\frak d$;

ii) $P-\wt P\in d^{1/2}$, $Q-\wt Q\in d^{1/2}$, and
$(\wt P-P)\,dQ=(\wt Q-Q)\,dP+O(d)$;

iii) $\wt W-W=(1/2)\langle P+\wt P,\wt Q-Q\rangle + O(d^{3/2})+c$, where
$c$ is some constant.
\end{DefinitioN}

\begin{DefinitioN} 
\rm
A Lagrangian asymptotic manifold $L$ in $\R^{2n}$ is a collection of the
following data: a closed subset $\Ga=\Ga_L\subset\R^{2n}$ (the {\it
support\/} of $L$) and a family $\{r_a\}_{a\in\mathcal{A}}$ of pairwise
consistent Lagrangian charts (an {\it atlas\/} of $L$) such that
$\Ga(r_a)$ is a relatively open subset in $\Ga$ for each $a\in\mathcal{A}$ and
$\cup_{a\in\mathcal{A}}\Ga(r_a)=\Ga$.
\end{DefinitioN}

One does not distinguish Lagrangian asymptotic manifolds with equivalent
atlases (two atlases are said to be {\it equivalent\/} if their union is
itself a valid atlas), and in what follows we assume that the atlas in
Definition~3.3 is maximal (i.e., is the union of all atlases in an
equivalence class).

If $(p,q)\in\Ga_{(r_a)}$, then we say that $r_a$ is a {\it chart\/} in a
neighborhood of the point $(p,q)$ on $L$, or that the chart $\Ga_a$
covers the point $(p,q)$.

\begin{DefinitioN} 
\rm
Let $I\subset\{1,\dots,n\}$. A Lagrangian chart $V=(U,d,P,Q,W)$ is said
to be $I$-{\it nonsingular\/} if $(Q_I(\a),P_{\ov I}(\a))=\a$. An
$I$-nonsingular chart with $I=\{1,\dots,n\}$ is merely said to be {\it
nonsingular\/} without mentioning $I$. The function $S_I(q_I,p_{\ov I})=
W(q_I,p_{\ov I}) - p_{\ov I}Q_{\ov I}(q_I,p_{\ov I})$ is called the
$I$-{\it phase\/} in $r$.
\end{DefinitioN}

\begin{LemmA} 
Let $L$ be a Lagrangian asymptotic manifold in $\R^{2n}$. Then each point
$(p,q)\in\Ga_L$ is covered by an $I$-nonsingular chart for some
$I\subset\{1,\dots,n\}$.
\end{LemmA}

The proof is immediate from the following lemma.
\medskip

\noindent{\bf Lemma (on local coordinates).}
{\it Let $V\subset\C^{2n}_{\xi,\eta}$ be a complex Lagrangian plane {\rm(}that
is, $\op{dim}V=n$ and $d\xi\wedge d\eta|_V=0${\rm)}. Then there exists a
subset $I\subset\{1,\dots,n\}$ such that $(\xi_I,\eta_{\ov I})$ is a
coordinate system on $V$ {\rm(}that is, the differentials $(d\xi_I|_V,
d\eta_{\ov I}|_V)$ are linearly independent{\rm)}.}
\medskip

The proof can be found in \cite{M2}, p.~369, and elsewhere.

The atlas consisting of $I$-nonsingular charts with various $I$ will be
called the {\it canonical covering\/}.

Let a point $(p_0,q_0)\in\Ga_L$ be covered by an $I$-nonsingular chart
$r_I$ and a $K$-nonsingular chart $r_k$ for some
$I,K\subset\{1,\dots,n\}$. Let us write out the formula relating the
corresponding $I$-phase and $K$-phase $S_K$. By applying the
transformation $\ga_K$, we can reduce the problem to the case $\ov
K=\varnothing$. We denote the chart $r_K$ simply by $r$ and the phase
$S_K$ simply by $S$. In this notation,
\begin{equation}
S_I(q_I,p_{\ov I}) = \Bigg\{
S(q) -q_{\ov I}p_{\ov I} -\frac12 \Big\langle
p_{\ov I} - \frac{\pa S}{\pa q_{\ov I}},
\Big(\frac{\pa^2 S}{\pa q_{\ov I}
\pa q_{\ov I}}\Big)^{-1}\Big(p_{\ov I}
-\frac{\pa S}{\pa q_{\ov I}}\Big)\Big\rangle\Bigg\}_{q_{\ov I} =
q_{\ov I}(q_I,p_{\ov I})} +O(d_I^{3/2}),
\tag{3.3}
\end{equation}
where $q_{\ov I}=q_{\ov I}(q_I,p_{\ov I})$ is an arbitrary smooth mapping
such that $q_{\ov I}(q_{0I},q_{0\ov I})=q_{0\ov I}$ and
$p_{\ov I}-\pa S/\pa q_{\ov I}(q_I,q_{\ov I}(q_I,p_{\ov I}))
=O(d_I^{1/2})$.

\begin{DefinitioN} 
\rm
A Lagrangian asymptotic manifold $L$ is said to be {\it
positive\/}\footnote{We prefer this term to the term
``dissipative" used in \cite{VDM2} and some other papers. Maybe
``nonnegative" would be even a better choice, but we  use ``positive."}
if
for any $I\subset\{1,\dots,n\}$ and any $I$-nonsingular chart on $L$, the
function $\Im S_I(q_I,p_{\ov I})$ is equivalent to $d(q_I,p_{\ov I})$ in a
sufficiently small neighborhood of $\Ga_d$, that is, the {\it dissipativity
inequality\/}
\begin{equation}
cd(p_I,q_{\ov I})\leqslant \Im S_I(q_I,p_{\ov I})\leqslant Cd(p_I,q_{\ov I})
\tag{3.4}
\end{equation}
is valid with some positive constants $c$ and $C$ in a neighborhood of
each point of $\Ga_d$.
\end{DefinitioN}

\begin{LemmA} 
Let $(p_0,q_0)\in\Ga(r_I)\cap \Ga(r_k)$, where $r_I$ and $r_K$ are an
$I$- and a $K$-nonsingular chart on $L$, respectively. Then the
dissipativity inequality is valid for $\Im S_I$ in the chart $r_I$ in a
neighborhood of $(q_{0 I},p_{0\ov I})$ if and only if it is valid for
$\Im S_K$ in the chart $r_K$ in a neighborhood of $(q_{0K},p_{0\ov K})$.
\end{LemmA}

The proof can be found, say in \cite{M2}, p.~386, or \cite{VDM2}, p.~104;
however, later on in this paper we shall give an independent proof
based on a lemma that will also prove useful when we shall consider
canonical transformations.

\subsection{Lagrangian asymptotic manifolds as asymptotic manifolds:\\ local description}

Given a Lagrangian asymptotic manifold $L$ in the sense of
Definition~3.3, it is easy to interpret $L$ as an asymptotic manifold in
the sense of \S1 and \S2. Namely,
let a  Lagrangian
chart $r=(U,d,P,Q,W)$ be given. The quadruple $(U,d,P,Q)$ determines an
$n$-dimensional asymptotic submanifold $L=(\mathcal{D}, \mathcal{J})$ in
$\R^{2n}_{p,q}$ in the standard way (parametric local description, see
\S1.4): we set
\begin{align}
\wh{D}(p,q,\a) &= d(\a) + \|p-P(\a)\|^2 + \|q-Q(\a)\|^2,
\nonumber\\
\wh{\mathcal{J}} &= \wh{\mathcal{D}} + \{p_1-P_1(\a),\dots,p_n-P_n(\a),
q_1-Q_1(\a),\dots,q_n-Q_n(\a)\},
\tag{3.5}
\end{align}
where $\wh{\mathcal{D}}$ is the dissipation ideal generated by $D(p,q,\a)$;
then we find
$$
D(p,q)=\min_\a \wh D(p,q,\a),
$$
consider the dissipation
ideal $\mathcal{D}$   associated with $D$, and set
$$
\mathcal{J}=\{f(p,q,\a)\in\wh{\mathcal{J}}\mid f\text{ is independent of } \a\}.
$$

\begin{LemmA} 
{\rm(a)} The manifold $L$ is {\sl involutive\/} in the
sense that
$\{\mathcal{J}, \mathcal{J}\}\subset\mathcal{J}$,
where $\{\cdot\,, \cdot\}$ is the standard Poisson bracket corresponding to
the symplectic structure \thetag{3.1}.

{\rm (b)} Consistent Lagrangian charts determine the same asymptotic
submanifold on their intersection.
\end{LemmA}
\medskip

\noindent{\bf Remark.} Note that the converse of Lemma~3.8~(b) is not true, since
condition iii) in Definition~3.2 does not follow from i) and ii) (nor does
the very existence of a function $W$ satisfying condition iii) in
Definition~3.1 follow from conditions i) and ii) in that
definition).
\medskip

For this reason, we must retain the phase $W_I$, i.e., incorporate
it in the new definition of Lagrangian asymptotic manifold to be
devised; this is done in the next subsection.
\medskip

\noindent{\it Proof of Lemma}~3.8.
Let us prove (b). Let $r$ and $\wt r$ be two consistent charts.
Without loss of generality we can assume that $V=U$ and $\wt V=\wt U$.
It readily follows from condition ii) in Definition~3.2 that $\wh D(p,q,\a)$ and
$$
\wh{\wt D}(p,q,\a)=d(\a) +\|p-\wt P(\a)\|^2 +\|q-\wt Q(\a)\|^2
$$
are equivalent; hence, so are
$$
D(p,q)=\min_\a\wh D(p,q,\a)\quad\text{ and }\quad
\wt D(p,q)=\min_\a\wh{\wt D}(p,q,\a).
$$
The ideal
$
\wh{\!\wt{\mathcal{J}}} = \wh D + \{p_1-P_1(\a),\dots,
p_n-P_n(\a),q_1-Q_1(\a),\dots,q_n-Q_n(\a)\}
$
is obviously different from $\wh{\mathcal{J}}$; however, the ideal $\mathcal{J}$
coincides with the ideal
$$
\wt{\mathcal{J}}=\{f(p,q,\a)\in\wh{\!\wt{\mathcal{J}}}\mid
f \text{ is independent of } \a\}.
$$
Indeed, let $f(p,q)\in\mathcal{J}$. Then
$$
f(p,q) = A(\a,p,q)(p-P(\a)) + B(\a,p,q)(q-Q(\a)) + O(\wh D).
$$
Differentiating $f(p,q)$ with respect to $\a$ yields
$$
A\frac{\pa P}{\pa\a} + B\frac{\pa Q}{\pa\a} = O(\wh D^{1/2}).
$$
It follows from Lemma~3.6 that $\op{det}(\pa Q_I/\pa\a,\pa P_{\ov I}/\pa\a)\ne0$
for some $I\subset\{1,\dots,n\}$; by applying $\ga_I$
we can assume without loss of generality that $I=\{1,\dots,n\}$. Then
$A(\pa P/\pa Q)+B = O(\wh D^{1/2})$. Next,
$$
f(p,q) = A(p-\wt P) + B(q-\wt Q) + A(\wt P-p) +B(\wt Q-Q) + O(\wh D).
$$
It follows from Definition~3.1, iv) that $\pa P/\pa Q-{}^t\pa P/\pa
Q=O(d^{1/2})$ and from Definition~3.2 that
$$
\wt P-P = \frac{{}^t\pa P}{\pa Q}(\wt Q-Q) + O(d) =
\frac{\pa P}{\pa Q}(\wt Q-Q) +O(d).
$$
Thus,
$$
A(\wt P-P)+B(\wt Q-Q) = \Big(A\frac{\pa P}{\pa Q}+B\Big)(\wt Q-Q) =
O(\wh D),
$$
and we see that $f(p,q)\in\wt{\mathcal{J}}$.
By symmetry, $\wt{\mathcal{J}}\subset \mathcal{J}$,
so that $\mathcal{J}=\wt{\mathcal{J}}$, and item (b) is
proved. It follows from (b) that it suffices to verify (a)
for the case in which the chart $r$ is $I$-nonsingular for some
$I\subset\{1,\dots,n\}$ (and even for $I=\{1,\dots,n\}$).
\qed \medskip

In an $I$-nonsingular chart the manifold $L=(\mathcal{D},\mathcal{J})$ can be
described more explicitly as follows.

\begin{LemmA} 
Let $r=(U,d,P,Q,W)$ be an $I$-nonsingular chart. Then the corresponding
Lagrangian manifold $L=(\mathcal{D}, \mathcal{J})$ is given by

{\rm i)} the dissipation
$$
D(p,q) = d(q_I,p_{\ov I}) +
\Big\|p_I-\frac{\pa S_I}{\pa q_I}\Big\|^2 +
\Big\|q_{\ov I}+\frac{\pa S}{\pa p_{\ov I}}\Big\|^2;
$$

{\rm ii)} the ideal
$$
\mathcal{J}=\mathcal{D} +\Big\{p_I-\frac{\pa S_I}{\pa q_I},
q_{\ov I}+\frac{\pa S}{\pa p_{\ov I}}\Big\}.
$$
Here
$S_I(q_I,p_{\ov I})$
is the $I$-{\sl phase\/} in the $I$-nonsingular chart on $L$.
\end{LemmA}

The proof is by straightforward computation using Definition~3.1 and
Eq.~\thetag{3.5}.

We can now finish the proof of Lemma~3.8. Assuming $I=\{1,\dots,n\}$,
we have $\mathcal{J}=D+\{p-\pa S/\pa q\}$, and involutivity follows
readily, since for the Poisson bracket we have
$$
\Big\{p_i-\frac{\pa S}{\pa q_i},
p_s-\frac{\pa S}{\pa q_s} \Big\}= \frac{\pa^2 S}{\pa q_i \pa q_s} -
\frac{\pa^2 S}{\pa q_s \pa q_i}=0.
$$
Lemma~3.8 is proved.
\medskip

Thus, to any Lagrangian asymptotic manifold in the sense of Definition~3.3
we have assigned an asymptotic manifold in the sense of \S1. However,
the inverse correspondence is not clear as yet; to guarantee its
existence, we must first incorporate phases in the definition; this
will be done in the next subsection.

\subsection{Global definition}

We shall sometimes use the ``complex coordinates" $(z,\ov z)$ on
$\R^{2n}$, where $z=(z_1,\dots, z_n)$, $\ov z=(\ov z_1,\dots,\ov
z_n)$, and
$$
z_j = q_j-ip_j,\quad \ov z_j = q_j +ip_j,\quad j=1,\dots,n.
$$
Let $\mathcal{D}$ be a dissipation ideal in $C^\infty(\R^{2n})$, and let
$\Ga=\op{loc}(\mathcal{D})$ be the set of its zeros. Furthermore, let $U\subset
\R^{2n}$ be a sufficiently small neighborhood of $\Ga$, and let
$$
\pi:\wt U\to U
$$
be the universal covering over $U$. Then $\wt U$ is a simply connected
manifold. The mapping $\pi$ is a local diffeomorphism, and so we can
freely use the same coordinates in $U$ and in $\wt U$. Furthermore, the
ideal $\pi^*(\mathcal{D})$ is well defined in $C^\infty(\wt U)$; for
brevity, it will be denoted by the same letter $\mathcal{D}$.

\begin{DefinitioN} 
\rm
A {\it $z$-action\/} is an element $\Phi\in C^\infty(\wt U)/{\mathcal{D}}^{3/2}$
that satisfies the following three conditions.

i) Let $m_0\in\Ga$ be an arbitrary point, and let $\Phi_1(p,q)$ and
$\Phi_2(p,q)$ be two branches of $\Phi$ defined in a neighborhood of
$m_0$. Then
$$
\Phi_1(p,q) - \Phi_2(p,q) = \Phi_1(m_0)-\Phi_2(m_0)+O({\mathcal{D}}^{3/2})
$$
in a neighborhood of $m_0$. In other words, the values of $\Phi$ on any
two sheets of the covering $\pi$ differ by a constant modulo $O({\mathcal{D}}^{3/2})$.

ii) There exists a vector function
$$
Z^*(p,q) = (Z_1(p,q),\dots,Z_n(p,q))\in C^\infty(U)/\mathcal{D}
$$
such that
$$
d\Phi=\frac{1}{2i} Z^*\,dz +O(\mathcal{D})
$$
(note that $Z^*$ is a function on $U$ rather than on $\wt U$, which is not
surprising in view of condition i)).

iii) The function $Z^*$ satisfies the condition
$
\ov z_j - Z_j^* \in{\mathcal{D}}^{1/2}$, $ j=1,\dots,n$.
\end{DefinitioN}

Suppose that a dissipation ideal $\mathcal{D}$, a covering $\pi:\wt U\to U$,
and a $z$-action $\Phi$ are given. We shall now construct the
corresponding Lagrangian asymptotic manifold.

Set $\mathcal{J}=\mathcal{D}+\{\ov z_1 - Z_j^*,\dots,\ov z_n - Z_n^*\}$. Then $L=(\mathcal{D},
\mathcal{J})$ is obviously an asymptotic
manifold of dimension $n$.

Furthermore, $\{\mathcal{J}, \mathcal{J}\}\subset\mathcal{J}$ (this can easily be
proved by straightforward computation), and so in a neighborhood of
each point of $\Ga$ the functions $(q_I,p_{\ov I})$ for some
$I\subset\{1,\dots,n\}$ are coordinates on $L$. By $U_I$ we denote the
projection of this neighborhood on the coordinate plane
$(q_I,p_{\ov I})$.

Let us construct the corresponding Lagrangian charts. By applying the
transformation $\ga_I$, we can always assume that we are in a
nonsingular chart. Set
$$
Q(p,q) = \frac{Z^*+z}{2},\qquad P(p,q) = \frac{Z^*-z}{2}.
$$
Then
$Z^*=Q+iP$, $z=Q-iP$.
We define
\begin{equation}
W=\Phi+\frac{\langle P,Q\rangle}{2} - \frac{P^2+Q^2}{4i}.
\tag{3.6}
\end{equation}
Then straightforward computation shows that
\begin{equation}
dW=P\,dQ.
\tag{3.7}
\end{equation}
Furthermore, since $dQ_1,\dots,dQ_n$ are linearly independent, we can
complete them  by some differentials $dF_1,\dots,dF_n$ to form a
basis of differentials on $\R^{2n}$ in a neighborhood of $(p_0,q_0)$.
By differentiating \thetag{3.7}, we obtain
$$
dP=\mathcal{E}\,dQ +\mu\,dF,
$$
where $\mathcal{E}-{}^t\mathcal{E} = O({D}^{1/2})$, $\mu= O({D}^{1/2})$,
and $\mathcal{E}$ is independent of the choice of $F_1,\dots, F_n$ modulo $
O({D}^{1/2})$. Let
$$
p(q)=\op{arg}\min_p  D(p,q).
$$
We set
\begin{equation}
S(q)=\Big\{W+\langle P, q-Q\rangle +\frac12 \langle q-Q,
\mathcal{E}(q-Q)\rangle\Big\}\Big|_{p=p(q)}.
\tag{3.8}
\end{equation}

The corresponding formulas for $I\ne\{1,\dots,n\}$ read
\begin{align}
W_I&= W-P_{\ov I} Q_{\ov I},
\nonumber\\
S_I(q_I,p_{\ov I}) &= \{W_I +\langle P_I,q_I-Q_I\rangle -
\langle Q_{\ov I}, p_{\ov I}-P_{\ov I}\rangle
\nonumber\\
&\quad +\,\frac12 \left\langle
\binom{q_I-Q_I}{p_{\ov I}-P_{\ov I}},{\mathcal{E}}_I
\binom{q_I-Q_I}{p_{\ov I}-P_{\ov I}}\right\rangle
\bigg\}_{\substack{
p_I=p_I(q_I,p_{\ov I})\\
q_{\ov I}=q_{\ov I}(q_I,p_{\ov I})}},
\tag{3.9}
\end{align}
where the matrix function $\mathcal{E}_I$ is defined from the condition
$$
d(P_I,Q_{\ov I}) = \mathcal{E}_I d(Q_I,-P_{\ov I}) + O(D^{1/2})
$$
and
$$
(p_I(q_I,p_{\ov I}),q_{\ov I}(q_I,p_{\ov I})) = \op{arg}
\min_{p_I,q_{\ov I}}D(p,q).
$$

Furthermore, we set
\begin{equation}
d_I(q_I,p_{\ov I}) = \min_{P_I,q_{\ov I}}D(p,q).
\tag{3.10}
\end{equation}

\begin{LemmA} 
{\rm(a)} The function $S_I(q_I,p_{\ov I})$ does not depend modulo
$O(d_I^{3/2})$ on the choice of the representative of $\Phi\in
C^\infty(\wt U)/D^{3/2}$ in $C^\infty(\wt U)$.

{\rm(b)} The quintuple
$$
r_I = \Big(U_I,d_I,\Big(p_{\ov I},\frac{\pa S_I}{\pa q_I}\Big),
\Big(q_I,-\frac{\pa S_I}{\pa p_{\ov I}}\Big),S_I\Big)
$$
is an $I$-nonsingular Lagrangian chart associated with the asymptotic
manifold $(\mathcal{D}, \mathcal{J})$.

{\rm(c)} All Lagrangian charts described in {\rm(b)} are pairwise
consistent.
\end{LemmA}

\noindent{\it Proof.}
(a) We can assume that $I=\{1,\dots,n\}$. Equation \thetag{3.6} for
$W$ can be rewritten as follows:
$$
W=\Phi+\frac{1}{8i}((Z^*)^2-z^2-2Z^*_{\mathcal{Z}}).
$$
Let $\wt\Phi=\Phi+O(D^{3/2})$. It follows that $P-\wt P=O(D)$, $Q-\wt
Q = O(D)$, and $\mathcal{E}-\wt{\mathcal{E}}=O(D^{1/2})$.

Furthermore,
\begin{align*}
\wt W-W &=\wt\Phi-\Phi+\frac{1}{8i}({\wt Z}^{*2} - {Z^*}^2 -2z({\wt
Z}^*-Z^*)) = \frac{1}{8i}({{\wt Z}^*}-Z^*)({\wt Z}^*+Z^*-2z)\\
&=\frac{1}{8i}({{\wt Q}^*}-Q)\times 2i(\wt P+P) = \frac12(\wt P+P)(\wt
Q-Q)=P(\wt  Q-Q)+O(D^{3/2}).
\end{align*}
Further, we obtain
\begin{align*}
S&= W+\langle P,q-Q\rangle +\frac12 \langle q-Q,\mathcal{E}(q-Q)\rangle,\\
\wt S&=\wt W+\langle \wt P,q-\wt Q\rangle +\frac12 \langle q-\wt
Q,\wt{\mathcal{E}}(q-\wt Q)\rangle,\\
\wt S-S&=\wt W- W+\langle P,Q-\wt Q\rangle  + O(D^{3/2}) =
O(D^{3/2}),
\end{align*}
and (a) is proved.

(b) Again we assume that $I=\{1,\dots,n\}$. Then what we need to prove
is that the functions $p-\pa S/\pa q$ generate the same ideal as $\ov
z-Z^*$. This can be proved by straightforward computation.

(c) This can be verified by straightforward computation. Lemma~3.11 is
proved.
\qed \medskip

To prove that the traditional description of positive Lagrangian
asymptotic manifolds is equivalent to that via the $z$-action, it remains
to explain how to reconstruct the $z$-action from the phases. The
answer is given by the following lemma.

\begin{LemmA} 
Let $L$ be a positive Lagrangian manifold, and let $S(q)$ be a
nonsingular phase on $L$. Then the function
\begin{align*}
\Phi(p,q) &= S(q) -\frac12 pq +\frac{q^2}{4i} - \frac{(\pa S/\pa q)^2}{4i}
\\
&\quad - \frac{1}{4i} \Big\langle p-\frac{\pa S}{\pa q},
\Big(1-i\frac{\pa^2S}{\pa q\pa q}\Big)^{-1}
\Big(1+i\frac{\pa^2S}{\pa q\pa q}\Big)
\Big(p-\frac{\pa S}{\pa q}\Big)\Big\rangle
\end{align*}
is the $z$-action on $L$.
\end{LemmA}

The proof is by straightforward computation.

Note that for positive Lagrangian manifolds the matrix $(1-i(\pa^2S/\pa q\pa
q))$ is always nonsingular, and  positivity is essential here. The
formulas for constructing $\Phi(p,q)$ from an $I$-nonsingular phase
$S_I(q_I,p_{\ov I})$ are obtained by applying $\ga_I$.

Let us now give the  independent proof (promised above) of the fact that
positivity is preserved in transition from one $I$-nonsingular chart
to another (Lemma~3.7). It suffices to consider the case in which
$I=\{1,\dots,n\}$ and $K=\varnothing$ (the variables $x_{\ov K}$ and
$p_{\ov K}$ can be regarded as parameters). Then, in view of the
transition formula \thetag{3.3}, Lemma~3.8 is a consequence of the
following general statement.

\begin{LemmA} 
Let $F(p,q)=F_1(p,q)+iF_2(p,q)$ be a smooth function satisfying the
conditions $F_2(p,q)\geqslant 0$, $F_2(p_0,q_0)=0$, $(\pa F/\pa
q)(p_0,q_0)=0$, and
$$
\op{det}\frac{\pa^2F}{\pa q\pa q}(p_0,q_0)\ne0.
$$
Also let
$$
D(p,q)=F_2(p,q) +\Big\|\frac{\pa F}{\pa q}(p,q)\Big\|^2.
$$
Then

{\rm(a)} the minimization problem
$$
D(p,q)\to\min_q \quad(\text{the minimum is taken over a small
neighborhood of }\,\, q_0)
$$
has a unique, smooth solution $q=q(p)$ for $p$ close to $p_0$, and
$q(p_0)=q_0$.

{\rm(b)} Let $d(p)=D(p,q(p))$, and set
\begin{equation}
\wt F(p) = \Bigg\{F(p,q) -\frac12 \Big\langle\frac{\pa F}{\pa
q}(p,q),\Big(\frac{\pa^2 F}{\pa q\pa q}(p,q)\Big)^{-1}
\frac{\pa F}{\pa
q}(p,q) \Big\rangle
\Bigg\}\Bigg|_{q=q(p)}.
\tag{3.11}
\end{equation}
Then there exist nonnegative constants $c$ and $C$ such that
\begin{equation}
cd(p) \leqslant \wt F_2(p)\leqslant C\,d(p),
\tag{3.12}
\end{equation}
where $\wt F_2(p)=\Im \wt F(p)$ is the imaginary part of $\wt F(p)$.
\end{LemmA}

The proof is given in the appendix.

Lemma~3.8 follows from Lemma~3.13 by setting
$F(p,q)=S(q)-pq$.

\subsection{Volume forms and the quantization condition}

As we established in \S3.3, a positive Lagrangian asymptotic manifold
is given by the following data: a closed subset $\Ga\subset\R^{2n}$, a
dissipation ideal $\mathcal{D}$ with $\Ga_{\mathcal{D}}=\Ga$, the universal
covering $\pi:\wt U\to U$ over a small neighborhood of $\Ga$, and a
$z$-action $\Phi(p,q)$ defined on $\wt U$. These data uniquely determine
the Lagrangian manifold $L=(\mathcal{D}, \mathcal{J})$ itself, and the
$I$-nonsingular phases in the charts of the canonical cover are given
by formulas \thetag{3.9}. Note that the canonical cover is in fact a
cover of $\wt U$ rather than of $U$; that is, the associated objects
($I$-nonsingular phases) depend on the choice of the sheet of $\wt U$.

We assume that a volume form $\mu$ is given on $L$. Since the form
$dz_1\wedge\cdots\wedge dz_n$ determines a nonzero element in
$\La^n(L)$, we can specify $\mu$ by choosing a fixed function $a(p,q)\in
\mathcal{C}^\infty(\R^{2n})$ such that
$$
\mu=i^*(a(q,p)\,dz_1\wedge\cdots\wedge dz_n).
$$
Of course only the class of $a(p,q)$ in ${\mathcal{C}}^\infty_{(1)}(\R^{2n})
= \mathcal{C}^\infty(\R^{2n})/\mathcal{D}^{1/2}$ is of
interest.

 We shall assume that the function $a(p,q)$ is defined on $\wt U$
rather than on $U$ (that is, the measure is defined on the universal
covering over $L$ rather than on $L$ itself).

Let $(p_0,q_0)\in\Ga$, let $V\subset U$ be a connected simply
connected neighborhood of $(p_0,q_0)$, and let $V_1$ and $V_2$ be two
connected components of $\pi^{-1}(V)\subset U$. By the definition of
the $z$-action, we have
\begin{equation}
\Phi_1(p,q)-\Phi_2(p,q) = \Phi_1(p_0,q_0)-\Phi_2(p_0,q_0)+O(D^{3/2}),
\tag{3.13}
\end{equation}
where $\Phi_i=\Phi|_{V_i}$, $i=1,2$.

Let $a_i(p,q)=a(p,q)|_{V_i}$, $i=1,2$.

Since $\wt U$ is simply connected, the expression
\begin{equation}
\op{Var}\op{ln}a = \op{ln}a_2-\op{ln}a_1 = \op{ln} (a_2/a_1)
\tag{3.14}
\end{equation}
is well defined in $V$. Indeed, let us arbitrarily choose the branch of
$\op{ln}a_1$; then the branch of $\op{ln}a_2$ is uniquely determined
by the condition that $\op{ln}a$ be continuous on $\wt U$. The
arbitrary multiple of $2\pi$ cancels in \thetag{3.14}, and
$\op{Var}\op{ln}a$ is well defined.

\begin{DefinitioN} 
\rm
A Lagrangian asymptotic manifold with $z$-action $\Phi$ and
measure $\mu=i^*a\,dz_1\wedge\cdots\wedge dz_n$ is said to satisfy the
quantization condition if for any $(p_0,q_0)\in\Ga$ and any two connected
components $V_1$ and $V_2$ of $\pi^{-1}(V)$, where $V$ is a small neighborhood
of $(p_0,q_0)$, we have
\begin{equation}
\Phi_1(p_0,q_0)-\Phi_2(p_0,q_0)+\frac{i}{2}(\op{Var}\op{ln}a)(p,q) =
O(D^{1/2})+2\pi l,
\tag{3.15}
\end{equation}
where $l\in\Z$ is an arbitrary integer.
\end{DefinitioN}

Condition \thetag{3.15} can be interpreted in two different ways.

First, we can regard it as a condition imposed on the admissible values of
$h$, which selects a sequence $h_l\to 0$.

Alternatively, if the Lagrangian manifold depends on parameters such as
energy, condition \thetag{3.15} selects admissible values of these parameters.

The following lemma is obvious.

\begin{LemmA} 
Suppose that $\Ga$ is arcwise connected; and let $\ga_1,\dots,\ga_s$ be a
fundamental system  of cycles on $\Ga$. The quantization condition
\thetag{3.15} is satisfied if and only if
\begin{enumerate}
\item[{\rm(a)}] $a_2/a_1$ is constant modulo $O(D^{1/2})$ for any branches
$a_1$ and $a_2$ of $a$;
\item[{\rm(b)}]
\begin{equation}
\op{Var}_{\ga_i}\Big[\frac{1}{h}\Phi+\frac{i}{2}\op{ln}a\Big]\in 2\pi\Z,\quad
i-1,\dots,s,
\tag{3.16}
\end{equation}
where $\op{Var}_\ga f$ is the variation of a function $f:\wt U\to\C$ along a
lift of a closed path $\ga\subset U$.
\end{enumerate}
\end{LemmA}

Condition \thetag{3.16} will also be referred to as the {\it quantization
condition\/}.
\medskip

\noindent{\bf Remark.}
For each $i$ Eq.~\thetag{3.16} gives infinitely many conditions, since
$\op{Var}_{\ga_i}$ may depend on the choice of the lift of $\ga$. However, if
$\Ga$ itself is a submanifold (necessarily isotropic), as is the case in
\cite{M3}, then
\begin{equation}
\op{Var}_{\ga_i}\Phi=\oint_{\ga_i} p\,dq
\tag{3.17}
\end{equation}
and does not depend on the choice of the lift; and furthermore, if $\Ga$ is a
closed trajectory of a Hamiltonian vector field and the measure $\mu$ is
invariant with respect to that field, then $\op{Var}_{\ga_i}(\op{ln}a)$ is also
independent of the lift and can be expressed via the Floquet exponents for the
variational system along this trajectory.
\medskip

\subsection{Positive canonical transformations}

Consider the space $\R^{4n}=\R^{2n}_{(p,q)}\oplus \R^{2n}_{(\xi,x)}$ equipped
with the symplectic form
$$
\Om^2=dp\wedge dq - d\xi\wedge dx.
$$
Let $\La=(\De,\mathcal{M})$ be a positive Lagrangian manifold in $\R^{4n}$ with
$z$-action $\Psi$ and suppose that $\La$ is ``diffeomorphically projected" on
$\R^{2n}_{(p,q)}$ and $\R^{2n}_{(\xi,x)}$ in the following sense:
\begin{enumerate}
\item[{(a)}] $\Ga$ is simply connected;
\item[{(b)}] the projections of $\Ga_\La$ on $\R^{2n}_{(p,q)}$ and on
$\R^{2n}_{(\xi,x)}$ are homeomorphisms onto their images;
\item[{(c)}] $(p,q)$ and $(\xi,x)$ are coordinate systems on $L$.
\end{enumerate}

\begin{DefinitioN} 
\rm
(i) The pair $g=(\La,\Psi)$ is called a {\it positive canonical
transformation\/} from $\R^{2n}_{(\xi,x)}$ to $\R^{2n}_{(p,q)}$. Formally,
we write  $g: \R^{2n}_{(\xi,x)}\to \R^{2n}_{\xi,x)}$.
\end{DefinitioN}

Let $L=(\mathcal{D}, \mathcal{J})$ be a positive Lagrangian manifold in
$\R^{2n}_{(\xi,x)}$ with $z$-action $\Phi$. Set
$$
g[\mathcal{D}](p,q)=\min_{x,\xi}(D(x,\xi)+\dt(x,\xi,p,q)),
$$
where $D$ and $\dt$ are some dissipations associated with $\mathcal{D}$ and
$\De$, respectively, and
$$
g[\mathcal{J}] = \{f(x,\xi,p,q)\in\mathcal{M}+\mathcal{J}\mid f\,\,\text{ is independent of
}\,\,(x,\xi)\}.
$$
Let $g[\mathcal{D}]$ be the dissipation ideal corresponding to the function $g[D]$.

\begin{LemmA} 
$g[L]=(g[\mathcal{D}], g[\mathcal{J}])$ is a positive Lagrangian manifold with $z$-action
\begin{equation}
g[\Phi](q,p) =
\Big\{F(x,\xi,p,q)-\frac12\Big\langle\frac{\pa F}{\pa(x,\xi)},
\frac{\pa^2 F}{\pa(x,\xi)\pa(x,\xi)}\frac{\pa F}{\pa(x,\xi)}\Big\rangle\Big\}
\Bigg|_{x=x(q,p),\,\xi=\xi(q,p)},
\tag{3.18}
\end{equation}
where
\begin{equation}
F(x,\xi,q,p)=\Phi(x,\xi) +\Psi(x,\xi,p,q)
\tag{3.19}
\end{equation}
and
\begin{equation}
(x(q,p),\xi(q,p)) = \op{arg}\min_{(x,\xi)}\mathcal{D}(x,\xi) +\De(x,\xi,p,q).
\tag{3.20}
\end{equation}
\end{LemmA}
\medskip

\noindent{\it Proof}.
The proof is purely technical, and we omit lengthy calculations; the only point
worth nothing is that positivity is preserved. To avoid using too many
subscripts, let us consider the particular case in which $L$ is covered by a
$\varnothing$-nonsingular chart and hence determined by a phase $S(\xi)$,
while $\La$ is determined by a phase $S_1(\xi,q)$. Then the nonsingular
phase $S_2(x)$ for $g(L)$ can be obtained as follows:
\begin{equation}
S_2(q)=\Big\{F(q,\xi) - \frac12 \Big\langle
F_\xi(q,\xi),\Big(\frac{\pa^2F}{\pa\xi\pa\xi}\Big)^{-1}F_p(q,\xi)
\Big\rangle\Big\}_{\xi=\xi(q)},
\tag{3.21}
\end{equation}
where $F(q,\xi)=S_1(\xi,q)+S(\xi)$ and
$$
\xi(q) = \op{arg}\min_{\xi,x}
\Big\{\Im S_1(\xi,q) +\Im S(q) +\Big\|x+\frac{\pa S(\xi)}{\pa\xi}\Big\|^2 +
\Big\|x-\frac{\pa S_1}{\pa\xi}(\xi,q)\Big\|^2
\Big\}.
$$
The dissipation on $g(L)$ in the nonsingular chart is
$$
d(q) = \min_{\xi,x} \Big\{\Im S_1(\xi,q) +\Im S(q) +\Big\|x+\frac{\pa
S(\xi)}{\pa\xi}\Big\|^2 + \Big\|x-\frac{\pa S_1}{\pa\xi}(\xi,q)\Big\|^2
\Big\}.
$$
It is easy to see that $d(q)$ is equivalent to
\begin{align*}
d_1(q) &= \min \Big\{\Im S_1 +\Im S_2 +
\Big\|\Re\frac{\pa S_1}{\pa\xi}+\Re\frac{\pa S}{\pa\xi}
\Big\|^2 +\Big\|\Im\frac{\pa S_1}{\pa\xi}
\Big\|^2 +\Big\|\Im\frac{\pa S}{\pa\xi}
\Big\|^2
\Big\}\\
&\simeq
\min \Big\{\Im (S_1 +S) +
\Big\|\frac{\pa (S_1+S)}{\pa\xi}
\Big\|^2
\Big\}
\end{align*}
(the last equivalence is due to the fact that $\Im S_1\geqslant 0$ and $\Im
S_2\geqslant 0$).

It remains to apply Lemma 3.13.
\qed \medskip

\begin{LemmA} 
Positive canonical transformations preserve quantization conditions.
\end{LemmA}

\noindent{\it Proof}. This is obvious, since the $z$-action on each sheet undergoes the
same additive correction, and the density of the volume form undergoes the same
multiplicative correction. \qed \medskip

\section{The canonical operator}

In this section we construct the first-approximation canonical operator on a
quantized Lagrangian asymptotic manifold with volume form. This material is
quite traditional (e.g., see \cite{M2, M3, MN, VDM2, MSS1, Ku2}), and we are
rather brief on the subject; our main goal is to relate the traditional
construction to the new definition of Lagrangian asymptotic manifold given
in~\S3.

\subsection{Original objects}

We assume that a positive asymptotic Lagrangian manifold with $z$-action and
with a volume form is given. That is, we have the following collection of
objects:
\begin{enumerate}
\item[{a)}] a closed subset $\Ga\subset \R^{2n}$, assumed to be arcwise
connected;
\item[{b)}] a dissipation ideal $\mathcal{D}\subset \mathcal{C}^\infty(\R^{2n})$ with
$\Ga_{\mathcal{D}}=\Ga$;
\item[{c)}] a small tubular neighborhood $U\supset\Ga$ and the universal
covering $\pi:\wt U\to U$;
\item[{d)}] a $z$-action $\Phi\in C^\infty(\wt U)$ satisfying the conditions
of Definition 3.10;
\item[{e)}] the corresponding asymptotic manifold $L=(\mathcal{D}, \mathcal{J})$;
\item[{f)}] a volume form $\mu=i^*(a(p,q)\,dz_1\wedge\cdots\wedge dz_n)$ on
$\pi^{-1}(L)$.
\end{enumerate}

We fix some branch of $\op{ln} a(p,q)$ on $\wt U$.
We assume that the quantization condition \thetag{3.15} is satisfied.

Furthermore, we choose a canonical cover $\wt U=\cup_j\wt U_j$, where each
$\wt U_j$ is a connected simply connected domain such that for some
$I=I(j)\subset \{1,\dots,n\}$ the functions $(q_I,p_{\ov I})$ can be chosen
as coordinates on $L$ in $U_j=\pi(\wt U_j)$. Although it may well happen
that $I(j)=I(k)$ for some $j\ne k$, we shall use the notation $\wt U_I$,
$U_I$ instead of $\wt U_j$, $U_j$ with $I(j)=I$; this will not lead to
any misunderstanding.

For each canonical chart $\wt U_I$ the $I$-phase $S_I(q_I,p_{\ov I})$ is
defined in the projection of $U_I$ on the coordinate $(q_I,p_{\ov I})$-plane
by Eq.~\thetag{3.9} and the dissipation $d_I(q_I,p_{\ov I})$ by
Eq.~\thetag{3.10}.

Furthermore, the form $\mu|_{\wt U_I}$ can be rewritten in the coordinates
$(q_I,p_{\ov I})$ as follows:
$$
\mu=a_I(q_I,p_{\ov I})\,dq_I\wedge dp_{\ov I},
$$
where
\begin{equation}
a_I(q_I,p_{\ov I}) = (-i)^{|\ov I|} a(p,q)\big|_{p_I=p_I(q_I,p_{\ov I}),\,\,
q_{\ov I}=q_{\ov I}(q_I,p_{\ov I})}
\op{det} \frac{\pa(q_I-iP_I,p_{\ov I}+iQ_{\ov I})}{\pa(q_I,p_{\ov I})} +
O(d_I^{1/2});
\tag{4.1}
\end{equation}
here
$$
(p_I(q_I,p_{\ov I}), q_{\ov I}(q_I,p_{\ov I})) = \op{arg}\min_{p_I,q_{\ov
I}}\mathcal{D}(p,q),\quad P_I=\frac{\pa S_I}{\pa q_I},\quad Q_{\ov I} = -
\frac{\pa S}{\pa p_{\ov I}}.
$$
We choose a continuous branch of $\op{ln}a_I(q_I,p_{\ov I})$ as follows.
Since $S_{I2}$ is nonnegative,
$$
B(q_I,p_{\ov I},r)=\op{det}\frac{\pa(q_I-i\tau P_I,p_{\ov I}+i\tau Q_{\ov
I})}{\pa(q_I,p_{\ov I})}\ne 0\quad\forall\tau\geqslant 0
$$
(e.g., see \cite{MSS1}). We have $b(q_I,p_{\ov I},0)=1$ and set $\op{ln}
B(q_I,p_{\ov I},0)=0$. Then, by continuity, $\op{ln}b$ is uniquely
determined for all $\tau>0$. We set
\begin{equation}
\op{ln}
a_I(q_I,p_{\ov I}) = \op{ln} a(p,q)\big|_{p_I=p_I(q_I,p_{\ov I}),\,\,
q_{\ov I}=q_{\ov I}(q_I,p_{\ov I})} +
\op{ln} b(q_I,p_{\ov I},1) -i\frac{\pi}{2}|\ov I| +
O(d_I^{1/2}),
\tag{4.2}
\end{equation}
where $\op{ln} a(p,q)$ is the fixed branch of the logarithm.
Equation~\thetag{4.2} specifies $\op{ln}a_I(q_I,p_{\ov I})$ uniquely.

\subsection{Local canonical operator}

Let $\wt U_I$ be some canonical operator chart. The {\it local canonical
operator\/}
$$
\mathcal{K}_I:C^\infty(L,U_I)\to H_h(\R^n)
$$
acts from the space
$$
C_0^\infty(L,U_I)=\{\ph\in C^\infty(L)\mid \op{supp}\ph\,\,\text{ is a
compact subset in }\,\, U_I\}
$$
to the Fr\'echet space
\begin{align*}
H_h(\R^n) &= \bigcap_{k=0}^\infty H_h^k(\R^n),\\
H_h^k(\R^n) &=\{f(q,h),\,\,q\in\R^n,\,\,h\in(0,1]\mid
\sup_h\|(1-h^2\De+x^2)^{k/2}f\|_{L^2(\R^n)} <\infty\}
\end{align*}
(here $\De=\sum_{i=1}^n \pa^2/\pa q^2$ is the Laplace operator) according to
the formula
\begin{equation}
[\mathcal{K}_I\ph] (q) = \Big(\frac{i}{2\pi h}\Big)^{|\ov I|/2}
\int_{\R^{|\ov I|}} e^{(i/h)(S_I(q_I,p_{\ov I}+p_{\ov I}q_{\ov I})}
\ph_I(q_I,p_{\ov I})\sqrt{a_I(q_I,p_{\ov I})}\,dp_{\ov I},
\tag{4.3}
\end{equation}
where $\ph_I(q_I,p_{\ov I})$ is the $(q_I,p_{\ov I})$-coordinate
representative of $\ph$, i.e., $\ph(q,p)-\ph_I(q_I,p_{\ov I})\in\mathcal{J}$.

\begin{LemmA} 
The operator \thetag{4.3} is well defined as an operator $f$ from
$C^\infty(L,U_I)$ to $H_h(\R^n)/ h^{1/2}H_h(\R^n)$ {\rm (}i.e., the image of
$\mathcal{K}_I\ph$ in the quotient space does not depend on the ambiguity in the
choice of representatives of $\ph_I,a_I$, and $S_I${\rm)}.
\end{LemmA}

\subsection{Global canonical operator and the commutation theorem}

Recall that we assume the quantization condition \thetag{3.15} to be
satisfied. Then the following assertion is valid.

\begin{Theorem} 
For any $\ph\in C^\infty(L,U_I)\cap C^\infty(L,U_K)$ we have
$$
\mathcal{K}_I\ph=\mathcal{K}_K\ph\quad\text{in}\quad H_h(\R^n)/h^{1/2} H_h(\R^n).
$$
Let $\{e_I(p,q)\}$ be a partition of unity subordinate to the canonical cover.
\end{Theorem}

We can now introduce the canonical operator as the operator
$$
\mathcal{K}:C_0^\infty(L)\to H_h(\R^n)/h^{1/2}H_h(\R^n)
$$
given by the formula
\begin{equation}
\mathcal{K} \ph = \sum_I \mathcal{K}_I(e_I\ph).
\tag{4.4}
\end{equation}
Obviously, the canonical operator is independent of the choice of
the partition of unity (by Theorem~4.2). Similarly,  for any $\e>0$, we
can define the canonical operator
\begin{equation}
\mathcal{K}:C_{0(1)}^\infty(L)\to H_h(\R^n)/h^{1/2-\e}H_h(\R^n).
\tag{4.5}
\end{equation}

\begin{Theorem}  
{\bf (Commutation theorem)}.
{\rm(a)} Let $H(q,p)$ be an arbitrary symbol {\rm(}i.e., a function
satisfying the estimates
$$
\Big|\frac{\pa^{\a+\be}H}{\pa x^\a\pa p^\be}(q,p)\Big|\leqslant
C_{\a\be}(1+|q|+|p|)^m,
$$
where $m$ is independent of $\a$ and $\be$). Then
\begin{equation}
H\Big(\overset 2  x,-\overset 1 {\frac{\pa}{\pa x}} \Big)
\mathcal{K}\ph = \mathcal{K}[(i^*H)\ph],
\tag{4.6}
\end{equation}
where $i^*H$ is the restriction of $H$ on $L$, that is, the image of $H\in
\mathcal{C}^\infty(\R^{2n})$ under the natural projection
$\mathcal{C}^\infty(\R^{2n})\to \mathcal{C}^\infty(L)$.

{\rm(b)} Let, in addition, $i^*H=0$, and suppose that $L$ is strongly
invariant with respect to the Hamiltonian vector field $V(H)$. Then
$$
\frac{i}{h}H\Big(\overset 2 x,-\overset 1 {\frac{\pa}{\pa x}} \Big)
\mathcal{K}\ph
$$
is a well-defined element in $H_h(\R^n)/h^{1/2-\e}H_h(\R^n)$ for any $\e>0$,
and we have
\begin{equation}
\frac{i}{h}H\Big(\overset 2 x,-\overset 1 {\frac{\pa}{\pa x}} \Big)
\mathcal{K}\ph = \mathcal{K} P\ph\quad\text{in}\quad
H_h(\R^n)/h^{1/2-\e}H_h(\R^n),
\tag{4.7}
\end{equation}
where the {\sl transport operator\/} $P$ is given by
$$
P=V(H)-\frac12 i^*\Big(\sum_{j=1}^n \frac{\pa^2H}{\pa p_i \pa q_i}\Big)+\frac12
\frac{\mathcal{L}_{V(H)}\mu}{\mu}
$$
{\rm(}note that the Lie derivative $\mathcal{L}_{V(H)}\mu$ is well defined since
$L$ is strongly invariant with respect to $V(H)${\rm)}.
\end{Theorem}

\subsection{Canonical operator for equations\\ with operator-valued symbol}

Let $H:\R^{2n}_{(p,q)}\to \op{Op}(\mathcal{H})$ be an operator-valued Hamiltonian
(see \cite{M2}) such that $H(p,q)$ is a self-adjoint operator for each
$(p,q)\in\R^{2n}$, and let $\la(p,q)$ be an isolated eigenvalue of constant
multiplicity $s$. Then associated with $(p,q)\in\R^n$ is the corresponding
$s$-dimensional eigenspace $E(p,q)$ of $H(p,q)$, and we have a vector bundle
$E\to\R^{2n}_{(p,q)}$ whose fiber over $(p,q)$ is $E(p,q)$. Let a quantized
Lagrangian asymptotic manifold $L$ with $z$-action be given; then the bundle
$\mathcal{E}=i^*(E)$ over $L$ is well defined, where $i$ is the ``embedding"
$L\hookrightarrow \R^{2n}$, and we can define the canonical operator
\begin{equation}
\mathcal{K}_{\mathcal{E}}:C_0^\infty(L;\mathcal{E})\to H_h(\R^n, \mathcal{H})
\tag{4.8}
\end{equation}
acting from the space of sections of $\mathcal{E}$ over $L$ into the Fr\'echet
space $H_h(\R^n_q, \mathcal{H})$ of $\mathcal{H}$-valued functions on $\R^n_q$ with the
topology defined by the family of norms
\begin{equation}
\|f\|_l=\|(1-h^2\La+q^2)^{l/2}f\|_{L^2(\R^n_q,\mathcal{H})}, \quad
l=0,1,2,\dots,
\tag{4.9}
\end{equation}
as follows. We define the local canonical operator by formula \thetag{4.3},
where $\ph_I$ now takes values in $\mathcal{H}$, and the global canonical operator
is defined by formula \thetag{4.4}.
Furthermore, we may well consider the canonical operator
$$
\mathcal{K}_{\mathcal{H}}:C_0^\infty(L,\mathcal{H})\to H_h(\R^q_n,\mathcal{H})
$$
acting from the space of arbitrary smooth compactly supported functions with
values in $\mathcal{H}$ by the same formulas. Then $\mathcal{K}_{\mathcal{E}}$ is the
restriction of $\mathcal{K}_{\mathcal{H}}$ to $C_0^\infty(L,\mathcal{E})$.

Lemma~4.1, Theorem 4.2, and Theorem 4.3 (a) remain valid in this situation
without any modifications. However, instead of Theorem~4.3~(b) we have the
following assertion.

\begin{Theorem} 
Suppose that the Lagrangian asymptotic manifold $L$ is strongly invariant with
respect to the Hamiltonian vector field $V(\la)$ corresponding to the
eigenvalue $\la$ and that $i^*\la=0$. Then for any $\ph\in C_0^\infty(L;\mathcal{E})$
the function $(i/h)\mathcal{K}_{\mathcal{E}}\ph$ is a well-defined element of the
quotient space
\begin{equation}
H_h(\R^n_q,\mathcal{H})/\mathcal{B},
\tag{4.10}
\end{equation}
where
$$
\mathcal{B}=\{\psi\in H_h(\R^n_q,\mathcal{H})\mid \psi
=\mathcal{K}_{\mathcal{H}}(i^*H)\eta\,\,\text{ for some }\,\, \eta\in C_0^\infty(L,\mathcal{H})\}.
$$
In the quotient space \thetag{4.10} we have
$$
\frac{i}{h}\mathcal{K}_{\mathcal{E}}\ph = \mathcal{K}_{\mathcal{E}}\mathcal{P} \ph,
$$
where the transport operator $\mathcal{P}$ has the form
$$
\mathcal{P}=\nabla_{V(\la)} -\frac12 i^*\Big(\sum_{j=1}^n \frac{\pa^2\la}{\pa
p_j \pa q_j}\Big) + M +\frac{\mathcal{L}_{V(H)}\mu}{\mu}.
$$
Here $\nabla_{V(\la)}$ is the covariant derivative along $V(\la)$ with respect
to the Levi-Civit\`a connection $\pa$ on $\mathcal{E}$ associated with the
operator of orthogonal projection onto the fibers of $\mathcal{E}$ in $\mathcal{H}$,
and the homomorphism $M:\mathcal{E}\to\mathcal{E}$ has the form $M=\langle dF,
V(\la)\rangle$ for some other homomorphism $F:\mathcal{E}\to\mathcal{E}$ {\rm(}cf.
\cite{DT}{\rm)}. In a local frame $\chi_1(p,q),\dots,\chi_s(p,q)$ in $E$
{\rm(}and hence in $\mathcal{E}${\rm)} we have
\begin{align}
(\nabla_{V(\la)})_{\mu\nu} &= V(\la)\dt_{\mu\nu} + (\chi_\nu, \dot\chi_\mu)
\qquad (\text{here }\,\,\dot\chi_\mu=V(\la)\chi_\mu),
\nonumber\\
M_{\mu\nu} &= \sum_{i=1}^n\Big(\chi_\nu, \Big(\frac{\pa H}{\pa
p_i}-\frac{\pa\la}{\pa p_i}\Big)\frac{\pa\chi_\mu}{\pa x_i}\Big),\quad
\nu,\mu=1,\dots,l.
\tag{4.11}
\end{align}
\end{Theorem}

Since $M$ is a total derivative, it does not contribute to the spectrum of
the transport operator. However, the term $(\chi_\nu, \dot\chi_\mu)$ does
contribute; its contribution along closed trajectories is an element of the
holonomy group of the connection $\dt$ and is known as {\it Berry's phase\/}
\cite{B, Si}. If the connection is flat, Berry's phase can be
incorporated in the quantization conditions in an obvious way.

\section{Canonical operator modulo
${\mathbf0 }{(}{\mathbf h}^{\boldsymbol{\infty}}{)}$}

\subsection{Asymptotic functions}

Here we introduce asymptotic (with respect to a small positive parameter $h$)
functions of the two following kinds:
\begin{enumerate}
\item[{i)}] $h,\mathcal{D}$-asymptotic functions on an asymptotic manifold; these
functions may be regarded as meromorphic in $h$ if taken modulo $h^N$,
$N\in\N$;
\item[{ii)}] $h$-asymptotic functions on $\R^n$ that may have an
``essential singularity" in $h$; in fact, we are interested in rapidly
oscillating $h$-asymptotic functions.
\end{enumerate}

First we recall some definitions. Let $(E_n)_{n\in\Z}$ be a filtration of a
vector space $E$. Then for any $e\in E$ we denote
$$
\op{ord}e=\sup\{n\in \Z \mid  e\in E_n\}.
$$

\begin{DefinitioN} 
\rm
An {\it asymptotic series\/} (with respect to
the filtration $(E_n)$) is a formal sum $\sum_{j=0}^\infty e_j$, where
$e_j\in E$ satisfy the condition $\lim \op{ord}e_j=+\infty$. Two asymptotic
series
$$
e'=\sum_{j=0}^\infty e_j'\,\,\,\text{ and }\,\,\,e''=\sum_{j=0}^\infty e_j''
$$
are
said to be {\it equivalent\/},
$e'\sim e''$, if
$$
\lim_{N\to\infty} \op{ord}
\sum_{j=0}^N(e_j'-e_j'')=+\infty.
$$
\end{DefinitioN}

\begin{DefinitioN} 
\rm
Let $A(E)$ be the set of all asymptotic series with respect to a given
filtration of $E$.  Elements of the quotient set $A(E)/\sim$ are called
{\it asymptotic elements\/} of $E$. If $E$ is a space of functions, then
asymptotic elements will be called {\it asymptotic functions\/}.
\end{DefinitioN}

Given $x_0\in R^n$, let $E(x_0)$ be the space of germs at $x_0$ of
$C^\infty$-smooth complex-valued functions smoothly depending on a {\it
strictly\/} positive small parameter $h$, and define a filtration of
$E(x_0)$ by setting
\begin{equation}
E_m(x_0) = \{f\in E(x_0)\mid f^{(\a)}=O(h^{m-k}) \text{ for any partial
derivative $f^\a$ of order $|\a|=k$}\}.
\tag{5.1}
\end{equation}
For example, $e^{(i/h)S}h^m u \in E_m(x_0)$ whenever $S$ and $u$ are germs
at $x_0$ of smooth functions independent of $h$ and $\Im S\geqslant 0$.

\begin{DefinitioN} 
\rm
An $h$-{\it asymptotic function\/} is a section of the sheaf $\mathcal{A}$ over
$R^n$ such that the stalk ${\mathcal{A}}_{x_0}$ consists of all the asymptotic
elements with respect to the filtration \thetag{5.1}.
\end{DefinitioN}

Let $L=(\mathcal{D}, \mathcal{J})$ be an $\infty$-asymptotic submanifold in some
manifold $M$. We define the space $E$ as the set of functions $f\in
C^\infty(L)\otimes C^\infty((0,\infty))$ satisfying the condition $h^k
f(m,h)\in C^\infty(L)\otimes C^\infty([0,\infty))$ for some $k\in \Z$. Let
us consider the following filtration of $E$:
\begin{align}
E_s = \{
&e\in E\mid e(m,h) =\sum_j e_j(m) h^j + h^k \wt e(m,h)
\nonumber\\
&\text{for some }\,\, e_j\in{\mathcal{D}}^{s/2-j},\,\, k\geqslant s/2,\,\,\wt e
\in C^\infty(L)\otimes C^\infty([0,\infty)) \};
\tag{5.2}
\end{align}
in \thetag{5.2} we assume that $j$ runs over a finite subset of $\Z$ (this
subset depends on $e$).

The reason for using the filtration \thetag{5.2} is clear from the
following lemma.

\begin{LemmA} 
Let $f\in E$. Then the following two relations are equivalent:
\begin{enumerate}
\item[{\rm i)}] $f\in E_s$;
\item[{\rm ii)}] in the vicinity of any point $m_0\in\Ga_{\mathcal{D}}$ the
inequality
$$
|\exp(-D(m)/h)f(m,h)|\leqslant ch^{s/2}
$$
holds for some constant $c$ {\rm(}here $D$ is a dissipation associated with
$\mathcal{D}${\rm)}.
\end{enumerate}
\end{LemmA}

\noindent{\it Proof}.
The statement of the lemma is equivalent to saying that for any smooth
function $g(x,h)$, any dissipation $D(x)$ in a neighborhood of $x_0$, and
any $s\in\Z$ the two following relations are equivalent:

1) $|\pa^j g/\pa h^j|_{h=0} = O(D^{(s/2)-j})$ for $j=0,1,2,\dots$;

2) $|\exp(-D(x)/h)g(x,h)|\leqslant ch^{s/2}$ for some constant $c$.
\qed \medskip

The proof of the latter statement can be found, for example, in
\cite{VDM1}.

\begin{DefinitioN} 
\rm
Asymptotic elements of $E$ corresponding to the filtration \thetag{5.2} will
be called ($h,\mathcal{D}$){\it-asymptotic functions\/} on $L$.
\end{DefinitioN}

\subsection{$\mathbf V$-objects on  asymptotic Lagrangian submanifolds\\
equipped with measure}

Suppose there is a positive asymptotic Lagrangian submanifold $L$ in
$\R^{2n}_{(p,q)}$ equipped
with a measure $\mu$, and let $\mathcal{D}$ be the associated dissipation ideal on
$\R^{2n}$. In this subsection we shall introduce some sheaf $\mathcal{V}$ over
the locus $\Ga$ of the ideal $\mathcal{D}$. Locally, in a chart, this sheaf can
be regarded as the ``bundle" of ($h,\mathcal{D}$)-asymptotic functions on $L$.

Let $m_0\in\Ga$ be an $I$-nonsingular point in $L$, $S_I$ and let $a_I$ be
well-defined modulo ${\mathcal{D}}^\infty$ functions which represent respectively
the branches of the $I$-phase and the density of the measure on the same
sheet of the simply connected covering of a neighborhood of $\Ga$. Then the
triple $T=(I,S_I,a_I)$ will be called a {\it trivializator\/} (for the
``bundle" $\mathcal{V}$ near $m_0$). Given a trivializator $T$, an
$\infty$-asymptotic manifold $L_\infty$ (with $L=L_\infty/\mathcal{D}$) is
locally defined, and we can identify the stalk ${\mathcal{V}}_{m_0}$ with the
space of germs at $m_0$ of ($h,\mathcal{D}$)-asymptotic functions on $L_\infty$.

Further, the latter can be identified with the space $L_I(m_0)$ of germs at
$(q_I^0,p^0_{\ov I})$ of ($h,\frak d_I$)-asymptotic functions on
$\R^n_{(q_I,p_{\ov I})}$, where $q^0=Q(m_0)$, $p^0=P(m_0)$, and $\frak d_I$
is the dissipation ideal on $\R^n_{(q_I,p_{\ov
I})}$ corresponding to the nonparametric local description of $L$:
if $\mathcal{D}$ is induced by a dissipation $D$, then $\mathcal{D}_I$ is associated with
$d(q_I,p_{\ov I})=\min_{q_{\ov I},p_{I}}D(p,q)$.
\medskip

\noindent{\bf Remark.}
Note that any manifold $M$ equipped with a dissipation ideal $\mathcal{D}$ can be
regarded as an $0$-codimensional $\infty$-asymptotic submanifold $(\mathcal{D},
\mathcal{D}^\infty)$ in itself.
\medskip

Now, to complete the definition of the sheaf $\mathcal{V}$, it is sufficient to
fix a certain family of gluing isomorphisms $V_{T,T'}^{m_0}: L_I(m_0)\to
L_{I'}(m_0)$ for each trivializator $T=(I,S_I,a_I)$,
$T'=(I',S_{I'}',a_{I'}')$ near $m_0$. Naturally, these isomorphisms are
assumed to satisfy the conditions
$$
V_{T,T}^{m_0} =\op{id},\quad V_{T',T''}^{m_0}\circ
V_{T,T'}^{m_0}=V_{T,T''}^{m_0}.
$$
We call $V_{T,T'}^{m_0}$ the {\it transition operators\/}. They will be
chosen later (see Eq.~\thetag{5.4}).

\begin{DefinitioN} 
\rm
A section of the sheaf $\mathcal{V}$ is called a $V$-{\it object\/} on $(L,\mu)$.
\end{DefinitioN}

\subsection{The canonical operator on $\mathbf V$-objects}

Suppose that we have a positive asymptotic Lagrangian submanifold $L$ in
$\R^{2n}_{(p,q)}$ with a measure $\mu$ together with
standard agreed-upon phase arguments (defined modulo $4\pi$) of its densities
relative to the Lagrangian coordinates $(q_I,p_{\ov I})$. Let $\mathcal{V}$ be
the corresponding sheaf of $V$-objects. For each $m\in\Ga$ and
any trivializator $T=(I,S_I,a_I)$ near $m$ we define the {\it precanonical
operator\/} $\wt{\mathcal{K}}_{m,T}:\mathcal{V}_m\to {\mathcal{A}}_{Q(m)}$
\begin{equation}
({\mathcal{K}}_{m,T}\ph)(q) = \Big(\frac{i}{2\pi h}\Big)^{|\ov I|}
\int \exp\Big(\frac{i}{h} S_I(q_I,p_{\ov I}) + p_{\ov I}q_{\ov I}\Big)
\sqrt{a_I(q_I,p_{\ov I})}e(p_{\ov I}) \ph(q_I,p_{\ov I})\, d p_{\ov I},
\tag{5.3}
\end{equation}
where $e$ is a smooth cutoff function equal to 1 near $P_{\ov I}(m)$
and to 0 near infinity, while $\ph(q_I,p_{\ov I})$ is an
($h,d$)-asymptotic function representing the germ $\ph\in L_I(m)={\mathcal{V}}_m$.
Note that the right-hand side of \thetag{5.3} does not depend on the
choice of $e$ modulo $O(h^\infty)$.

Assume now that the asymptotic Lagrangian submanifold $L$ satisfies the
quantization condition. Then the transition operators $V_{T,T'}^m$ in the
definition of the sheaf $\mathcal{V}$ can be uniquely chosen so that the
precanonical operator $\wt{\mathcal{K}}_{m,T}$ does not depend on $T$. In other
words, the transition operators are defined by the equation
\begin{align}
&\Big(\frac{i}{2\pi h}\Big)^{|\ov I|} \int \exp\Big(\frac{i}{h}
(S_I(q_I,p_{\ov I})+
 p_{\ov I}q_{\ov I})\Big)\sqrt{a_I(q_I,p_{\ov I})}\ph(q_I,p_{\ov
I})\,dp_{\ov I}
\nonumber\\
&\qquad
= \Big(\frac{i}{2\pi h}\Big)^{|\ov I'|} \int
\exp\Big(\frac{i}{h} (S_{I'}'(q_{I'},p_{\ov I'})+
 p_{\ov I'}q_{\ov I'})\Big)\sqrt{a_{I'}'(q_{I'},p_{\ov
I'})}(V_{T,T'}^m\ph)(q_{I'},p_{\ov I'})\,dp_{\ov I'}
\tag{5.4}
\end{align}
near $q=Q(m)$ for any ($h,d_I$)-asymptotic function $\ph$ supported near
$(Q_I(m),P_{\ov I}(m))$.

\begin{LemmA} 
There exist operators $V^m_{T,T'}$ such that \thetag{5.4} holds.
\end{LemmA}

\noindent{\it Proof}.
Applying the Fourier transform from $q_{\ov I'}$ to $p_{\ov
I'}$ and using the canonical transformation $\ga_{I'}$, we can assume
without loss of generality that $I'=\{1,2,\ldots,n\}$.

The statement is trivial in the case $\ov I=\varnothing$. Consider the case
$|\ov I|>0$.
\qed \medskip

Denote the left-hand side of \thetag{5.4} by $(K\ph)(q)$. Then $(K\ph)(q)$
can be expanded relative to the $h$-asymptotic filtration by using the
saddle-point method, say, in the form of the quantum bypassing focuses
operation introduced in \cite{M2}, Sec.~1 of Chap.~V. Using our notation, we
can formulate the result as follows. Consider the $I$-nonsingular
Lagrangian chart $r=(U_I,d_I,P',Q',W')$:
\begin{align*}
Q_I(\a) &=\a_I,\quad P'_{\ov I}(\a)=\a_{\ov I},\quad Q'_{\ov I}(\a) =-
\frac{\pa S_I(\a)}{\pa\a_{\ov I}},\\
P'_{ I}(\a) &= \frac{\pa S_I(\a)}{\pa\a_{ I}},\quad W(\a)=S_I(\a)+\a_{\ov I}Q'_{\ov
I}(\a).
\end{align*}

Let $y=(y_1,\dots,y_n)$ be nonsingular coordinates on $U_I$, i.e.,
$y-Q'=O(d^{1/2})$. Denote by $\Phi$ the $y$-phase on $U_I$ (cf.
\cite{VDM2}, \S1):
$$
\Phi=W+\frac{\pa W}{\pa Q'}(y-Q')+\frac12 \Big\langle
y-Q',\frac{\pa^2 W}{\pa Q'\pa Q'}(y-Q')
\Big\rangle.
$$
Then
$$
(K\ph)(y(\a)) =\exp \Big\{\frac{i}{h}\Phi(\a)\Big\}\sqrt{\wt a(\a)}
(v\ph)(\a)
$$
for some local (near $\a_0=(Q_I(m),P_{\ov I}(m))$) automorphism $v$ of the
sheaf of $(h,d_I)$-asympto\-tic functions on $U$, where $\wt a$ is the
density of the measure $a_I(\a)\,d\a_1\wedge\cdots\wedge d\a_n$ with
respect to the complex coordinates $Q'$ on $U$. Now, to complete the proof,
it is sufficient to verify that
\begin{align}
a\circ y-\wt a&= O(d_I^{1/2}),
\tag{5.5}\\
S\circ y-\Phi& =O(d_I^{3/2}).
\tag{5.6}
\end{align}
We consider the nonsingular Lagrangian chart $r''=\{U,d,P'',Q'',S\}$,
where $Q'(\be)=\be$, $P'(\be) = \pa S(\be)/\pa\be$. Here $U$ is a
neighborhood of $Q(m)$. Then, in some neighborhoods of the corresponding
images of the point $m$, the charts $r'$ and $r''$ are equivalent,
$y:\a\mapsto\be$ being a diffeomorphism identifying their domains. Hence
$d_I\circ y^{-1}$ is equivalent to $d$, and
$$
\Phi\circ y^{-1}-S\circ \op{id} =O(d^{3/2})
$$
as demonstrated in \cite{VDM2}, so that \thetag{5.6} holds.

Finally, we note that \thetag{5.5} holds, since the function $\wt a/a_I$ is
a representative of
$$
\frac{D(Q_I,P_{\ov I})}{DQ} \in C^\infty_{(1)},
$$
and the same
can be said about $(a\circ y)/a_I$. The lemma is proved.
\qed \medskip

\begin{DefinitioN} 
\rm
Given a quantized positive asymptotic Lagrangian submanifold $L\subset
\R_p^n\times\R^n_q$ with measure $\mu$, let us denote by $V_0$ the space
of $V$-objects with compact supports. We define the {\it canonical
operator\/} $\mathcal{K}:V_0\to As$, where $As$ is the space of $h$-asymptotic
functions on $R^n$ as follows: the germ $\mathcal{K}\ph$ at the point $x\in
Q(\Ga)$ is equal to $\sum_m{\mathcal{K}}_m \{\ph\}_m$, where $m$ runs over the
set  $\{m\in \op{supp}\ph\mid Q(m)=x\}$, and $\{\ph\}_m$ is the germ of
$\ph$ at $m$. We denote by ${\mathcal{K}}_m$ the precanonical operators, which
are independent of the choice of  trivializators under our definition of the
transition operators.
\end{DefinitioN}

\subsection{Commutation of the canonical operators with Hamiltonians}

Throughout this subsection, $\mathcal{H}(p,q,h)$ will be a real function smooth on
$\R^{2n}\times[0,\e]$ (the Hamiltonian function) satisfying
the following condition: there exists a positive integer $k$ such that for
any multiindex $\a=(\a_1,\dots,\a_{2n})$
$$
\Big(\frac{\pa}{\pa z}\Big)^\a \mathcal{H}(z,h)=O(|z|^k)\quad\text{as}\quad
z\to\infty;
$$
here $z=(p,q)\in\R^{2n}$. The pseudodifferential operator
$$
\mathcal{H}\Big(-\overset 1{ih\frac{\pa}{\pa x}}, \overset 2 x\Big)
$$
will be referred to as the {\it Hamiltonian\/} corresponding to the symbol
${\mathcal{H}}$. We assert that Hamiltonians are in agreement with $h$-asymptotic
filtration (for example, see \cite{M2} or \cite{VDM2}). Hence Hamiltonians
can and will be interpreted as linear operators in the space of
$h$-asymptotic functions.

We start from a version of the commutation formula for a complex bounded
exponential and a Hamiltonian \cite{VDM2}:
\begin{equation}
\mathcal{H}\Big(-\overset 1{ih\frac{\pa}{\pa x}}, \overset 2 x\Big)\circ
\exp\Big(\frac{i}{h} S(x)\Big) =\exp\Big(\frac{i}{h} S(x)\Big) \circ\wh{\mathcal{H}},
\tag{5.7}
\end{equation}
where $\wh{\mathcal{H}}$ is a linear operator in the space ${\mathcal{A}}_0$ of finite
($h,\mathcal{D}$)-asymptotic functions on $\R^n$, where $\mathcal{D}$ is the
dissipation ideal induced by the imaginary part of $S$.
\medskip

\noindent{\bf Example.}
Let $n=1$,
$$
\mathcal{H}(p,x) =\frac{p^2q^2}{2}+q.
$$
It is easy to verify that
\begin{align*}
\wh{\mathcal{H}} &=\exp\Big(-\frac{i}{h}S(x)\Big)\circ
\mathcal{H}\Big(-\overset 1{ih\frac{\pa}{\pa x}}, \overset 2 x\Big)\circ\exp
\Big(\frac{i}{h}S(x)\Big)\\
&= \mathcal{H}\Big(\overset 1{-ih\frac{\pa}{\pa x}+\frac{\pa S}{\pa x}},
\overset 2 x\Big)
= x^2 \Big(\frac{dS}{dx}\Big)^2 + x - ih x^2\Big(\frac{d^2S}{dx^2}
+2\frac{dS}{dx}\frac{d}{dx}\Big) - \frac{h^2}{2}x^2 \frac{d^2}{dx^2}.
\end{align*}
We see immediately that the operator $\wh{\mathcal{H}}$ does not decrease the order
of a function with respect to the ($h,\mathcal{D}$)-asymptotic filtration and
does not enlarge its support. Hence $\wh{\mathcal{H}}$ acts on $\mathcal{A}_0$.
\medskip

In general, the operator $\wh{\mathcal{H}}$ on the right-hand side of \thetag{5.7}
is described by an operator series of the form
\begin{equation}
\wh{\mathcal{H}} = \sum_{k=0}^\infty \sum_{k\leqslant|\a|\leqslant 2k}\,\,
\sum_{|\be|\leqslant k}(-ih)^k{\mathcal{H}}_{k\a\be}.
\tag{5.8}
\end{equation}
If the symbol $\mathcal{H}$ is independent of $h$,
then the operators ${\mathcal{H}}_{k\a\be}$ have the form
\begin{equation}
{\mathcal{H}}_{k\a\be} = \frac{\pa^{|\a|}\mathcal{H}}{\pa p^\a}
\Big(\frac{\pa S}{\pa x}, x\Big)P_{k\a\be}(S)\Big(\frac{\pa}{\pa x}\Big)^\be,
\tag{5.9}
\end{equation}
where $P_{k\a\be}$ are nonlinear differential operators and
$({\pa^{|\a|}\mathcal{H}}/{\pa p^\a})({\pa S}/{\pa x},x)$
stands for the Taylor expansion with respect to the imaginary part of $\pa
S/\pa x$. Note that $P_{k\a\be}$ can be calculated by using the fact that
they are independent of the symbol $\mathcal{H}$, choosing symbols in a
special way. Specifically, we have
\begin{align*}
P_{000}(S)&=1,\\
P_{1\a\be}(S) &= \begin{cases}
0\qquad &\text{for}\quad |\a|=1,\,\,\be=0\quad\text{or}\quad
|\a|=2,\,\,|\be|=1,
\\
\langle\a,\be\rangle\quad &\text{for}\quad |\a|=|\be|=1,
\\
\frac{1}{\a!}\frac{\pa^2S}{\pa x^\a}\quad &\text{for}\quad |\a|=2,\,\,|\be|=0.
\end{cases}
\end{align*}
Further, it is easy to see that $P_{k\a\be}=0$ for $2k<|\a|+|\be|$, and
$$
P_{k\a\be} = \begin{cases}
0\quad &\text{for}\quad |\a|=|\be|=k,\,\,\a\ne\be,\\
\frac{1}{\a!}\quad &\text{for}\quad \a=\be,\,\,|\a|=k.
\end{cases}
$$
Finally, in the case when $\mathcal{H}$ depends on $h$, we have
\begin{equation}
{\mathcal{H}}_{k,\a,\be} = \sum_{l=0}^k i^l {\mathcal{H}}^{(l)}_{k-l,\a,\be},
\tag{5.10}
\end{equation}
where
$$
{\mathcal{H}}^{(l)} = \frac{1}{l!}\frac{\pa^l\mathcal{H}}{\pa h^l}\Bigg|_{h=0}.
$$

Denote by $F_{x_{\ov I}\to\xi_{\ov I}}$ the $H^{-1}$-Fourier transformation
with respect to the $\ov I$th group of coordinates:
$$
F_{x_{\ov I}\to\xi_{\ov I}} u(x) = (2\pi ih)^{-|\ov I|/2}
\int \exp\Big\{-\frac{i}{h}\xi_{\ov I}x_{\ov I}\Big\}u(x)\,dx_{\ov I},
$$
and let $F_{\xi_{\ov I}\to x_{\ov I}}^{-1}$ denote its inverse. Then
$$
F_{x_{\ov I}\to\xi_{\ov I}}\circ
\mathcal{H}\Big(-\overset 1{ih\frac{\pa}{\pa x}}, \overset 2 x,h\Big)\circ
F_{\ov\xi_{I}\to x_{\ov I}}^{-1}=
\mathcal{H}\Big(\Big(-\overset 1{ih\frac{\pa}{\pa x_{\ov I}},\xi_{\ov
I}}\Big),  \Big(\overset 2{x_I,ih\frac{\pa}{\pa \xi_{\ov
I}}}\Big),h\Big).
$$
For the pseudodifferential operator on the right-hand side, there is a
commutation formula with a complex exponential, similar to
that for the Hamiltonian
$$
\mathcal{H}\Big(-\overset 1{ih\frac{\pa}{\pa x}}, \overset 2 x,h\Big).
$$
Thus, we obtain the following commutation formula for a Hamiltonian and the
composition of the multiplication operator by a complex exponential with the
Fourier transformation (see \cite{M2}, Sec.~2 in Chap. V)
\begin{equation}
\mathcal{H}\Big(-\overset 1{ih\frac{\pa}{\pa x}}, \overset 2 x\Big)
\circ F_{\xi_{\ov I}\to x_{\ov I}}^{-1}\circ e^{(i/h)S(x_I,\xi_{\ov I})} =
F_{\xi_{\ov I}\to x_{\ov I}}^{-1}\circ e^{(i/h)S(x_I,\xi_{\ov I})}
\wh{\mathcal{H}}_I,
\tag{5.11}
\end{equation}
where
\begin{align*}
\wh{\mathcal{H}}_I &= \sum_{k=0}^\infty \sum_{|\a|=k}^{2k} \sum_{|\be|=0}^k
\sum_{\ga\leqslant\be_{\ov I}}\,\,\sum_{|\dt|=0}^\infty \Bigg\{(-ih)^k
\frac{\be_{\ov I}!}{\ga!(\be_{\ov I}-\ga!)\dt!}
\Big(\Big(\frac{\pa}{\pa(q_{ I},p_{\ov
I})}\Big)^{\a+\dt}\Big(\frac{\pa}{\pa p_{\ov I}}\Big)^\ga\mathcal{H}\Big)
\\
&\quad\times
\Big(\Big(\frac{\pa S_1}{\pa x_I},\xi_{\ov I}\Big),
\Big(x_I,-\frac{\pa S_1}{\pa\xi_{\ov I}}\Big)\Big)
P^I_{k\a\be}(S)\Big(i\frac{\pa
S_2}{\pa x_I},-i\frac{\pa
S_2}{\pa p_{\ov
I}}\Big)^\dt\Big(\frac{\pa}{\pa
x_I}\Big)^{\be_I}\Big(\frac{\pa}{\pa\xi_{\ov I}}\Big)^{\be_{\ov I}-\ga} \Bigg\},
\end{align*}
where $S_1=\Re S$, $S_2=\Im S$, and $P_{k\a\be}^I$ are some nonlinear
differential operators independent of $\mathcal{H}$. (One can easily obtain the formula
for Hamiltonians depending on $h$.)

Now we are ready to commute a Hamiltonian with a canonical operator.

\begin{Theorem} 
Given a symbol $\mathcal{H}(p,q)$ and a quantized positive Lagrangian submanifold
$L$ equipped with a measure $\mu$, there is an operator $\mathcal{P}_{\mathcal{H}}$
acting on finite $V$-objects such that
$$
\mathcal{H}\Big(-\overset 1{ih\frac{\pa}{\pa x}}, \overset 2 x\Big)(\mathcal{K}\ph)(x)
= (\mathcal{K}{\mathcal{P}}_{\mathcal{H}}\ph)(x).
$$
\end{Theorem}

\noindent{\it Proof}.
Commute the Hamiltonian with a precanonical operator. Let $T=(I,S_I,a_I)$ be a
trivializator near a point $m\in\Ga$ and let $\ph(x_I,\xi_{\ov I})$ be the germ
of a ($h,d_I$)-asymptotic function at $(Q_I(m),P_{\ov I}(m))$. By
formula \thetag{5.11} we have
$$
\mathcal{H}\Big(-\overset 1{ih\frac{\pa}{\pa x}}, \overset 2 x\Big)
\wh{\mathcal{K}}_{m,T}\ph = F_{\xi_{\ov I}\to x_{\ov I}}^{-1}\exp\Big\{
\frac{i}{h}S_I(x_I,\xi_{\ov I})\Big\}
\wh{\mathcal{H}}_I \sqrt{a_I(x_I,\xi_{\ov I})}\ph(x_I,\xi_{\ov I}).
$$
It follows that
$$
\mathcal{H}\Big(-\overset 1{ih\frac{\pa}{\pa x}}, \overset 2 x\Big)
\wh{\mathcal{K}}_{m,T} = \wh{\mathcal{K}}_{m,T}{\mathcal{P}}_{m,T},
$$
where
$$
{\mathcal{P}}_{m,T} = \frac{1}{\sqrt{a_I}}\circ\wh{\mathcal{H}}_I\circ\sqrt{a_I}.
$$
If we interpret a precanonical operator as a local homomorphism of sheaves:
${\mathcal{V}}_m\to{\mathcal{A}}_{Q(m)}$, then it is a monomorphism independent of the
choice of a trivializator. Hence there is a morphism $\mathcal{P}$ of the sheaf
$\mathcal{V}$ to itself that, under the local trivialization of $\mathcal{V}$
determined by $T$, identifies $\mathcal{P}_m$ with ${\mathcal{P}}_{m,T}$. This
implies the required statement.

\subsection{The transport operator}

Here we consider a quantized positive asymptotic Lagrangian submanifold
with a measure $(L,\mu)$  strongly invariant with respect to the
Hamiltonian vector field $V_{\mathcal{H}}$ corresponding to a Hamiltonian function
$\mathcal{H}(p,q)$. In this case we call $P_{\mathcal{H}}=(i/h){\mathcal{P}}_{\mathcal{H}}$
the {\it transport operator\/}.

To describe the transport operator we need the following definition.

\begin{DefinitioN} 
\rm
\cite{VDM1} Let $L=(\mathcal{D},\mathcal{J})$ be an
$\infty$-asymptotic submanifold in a manifold $M$, and let the operator
series $\varkappa=\sum_{j=0}^\infty h^{t_j}A_j$, where $A_j$ are linear
operators from $C^\infty(L)$ to itself, represent an endomorphism of the
space of ($h,\mathcal{D}$)-asymptotic functions on $L$. We say $\ka$ is a {\it
perturbator\/} if the following conditions hold:
\begin{enumerate}
\item[{a)}] $h^{t_j}A_j$ does not decrease order relative to the
($h,\mathcal{D}$)-asymptotic filtration for any $j$;
\item[{b)}] if $t_j\leqslant 0$, then $h^{t_j}A_j$ increases the order.
\end{enumerate}
\end{DefinitioN}

Perturbators are valid in perturbation theory for equations with
($h,\mathcal{D}$)-asymptotic functions because of the following result.

\begin{PropositioN} 
{\rm(\cite{VDM1}, p.~84)}
Let $\ka$ be a perturbator
on $L=(\mathcal{D}, \mathcal{J})$. Then $\op{id}-\ka$ is an isomorphism of the space
of  ($h,\mathcal{D}$)-asymptotic functions on $L$, and its inverse is defined by
the operator series $\sum_{r=0}^\infty \ka^r$.
\end{PropositioN}

\begin{Theorem} 
The transport operator has the form
$$
P_{\mathcal{H}}=V_{\mathcal{H}}-\frac12 \op{tr} \frac{\pa^2 \mathcal{H}}{\pa p\pa
q}\Bigg|_L +\ka,
$$
where $V_{\mathcal{H}}$ is the Hamiltonian vector field associated
with $\mathcal{H}$, and $\ka$ is a perturbator depending on the choice of
a trivializator for $V$-objects.
\end{Theorem}

\noindent{\it Proof}.
Without loss of generality let us take some nonsingular
trivializator $(S,a)$. By using the commutation formula for a Hamiltonian
with an exponential, we obtain the following local expression for $P_{\mathcal{H}}$:
\begin{equation}
P_{\mathcal{H}} =\frac{1}{\sqrt a}
\Big\langle
\frac{\pa\mathcal{H}}{\pa p}\Big(\frac{\pa S}{\pa x},x\Big),
\frac{\pa}{\pa x}\Big\rangle\sqrt a + \frac12 \op{tr} \frac{\pa^2H}{\pa
p^2}\Big(\frac{\pa S}{\pa x},x\Big)\frac{\pa^2S}{\pa x^2}+\wt\ka,
\tag{5.12}
\end{equation}
where $\wt\ka$ is a perturbator. It is not difficult to show that the
invariance of $\mu$ with respect to $V_{\mathcal{H}}$ implies the following
result (similar to Liouville's theorem):
$$
V_{\mathcal{H}}\frac{1}{a} = \frac{1}{a}\op{tr} \Big(\frac{\pa^2\mathcal{H}}{\pa
p^2}\Big(\frac{\pa S}{\pa x},x\Big)\frac{\pa^2S}{\pa
x^2} +\frac{\pa^2\mathcal{H}}{\pa
p\pa q}\Big(\frac{\pa S}{\pa x},x\Big)\Big) +O(d^{1/2-\e}).
$$
Furthermore, the operator
$$
\Big\langle
\frac{\pa\mathcal{H}}{\pa p}\Big(\frac{\pa S}{\pa x},x\Big),
\frac{\pa}{\pa x}\Big\rangle
$$
represents $V_{\mathcal{H}}$ in our coordinates, and we obtain
\begin{equation}
\Big\langle
\frac{\pa\mathcal{H}}{\pa p}\Big(\frac{\pa S}{\pa x},x\Big),
\frac{\pa}{\pa x}\Big\rangle\frac{1}{\sqrt a} = -
\frac{1}{2\sqrt a}\op{tr}
\Big(\frac{\pa^2H}{\pa
p^2}\Big(\frac{\pa S}{\pa x},x\Big)\frac{\pa^S}{\pa x^2} +
\frac{\pa^2\mathcal{H}}{\pa p\pa q}
\Big(\frac{\pa S}{\pa x},x\Big)
\Big) + O(d^{1/2-\e}).
\tag{5.13}
\end{equation}
Substituting \thetag{5.13} into \thetag{5.12} completes the proof.
\qed \medskip

\section*{Appendix. Proof of Lemma 3.13}

 We write
$$
\mathcal{E}(q) =
\Big(\frac{\pa^2 F(p,q)}{\pa q_i \pa q_j}\Big)^n_{i,j=1}.
$$
Set
$$
D_\mu(p,q)=F_2(p,q) +\frac{\mu}{2}\Big\|\frac{\pa F}{\pa q}(p,q)\Big\|^2
$$
(where $\mu>0$ will be chosen later) and consider the problem
\begin{equation}
D_\mu(p,q)\to\min_q;
\tag{A.1}
\end{equation}
also, set
\begin{equation}
{\wt F}_\mu(p) = \Bigg\{F(p,q) -\frac12 \Big\langle\frac{\pa F}{\pa
q}(p,q),\Big(\frac{\pa^2 F}{\pa q\pa q}(p,q)\Big)^{-1}
\frac{\pa F}{\pa
q}(p,q) \Big\rangle
\Bigg\}\Bigg|_{q=q(p)},
\tag{A.2}
\end{equation}
where $q=q(p)$ is the solution to the minimization problem \thetag{A.1}. For
$\mu=2$ we obtain the functions \thetag{3.10}, \thetag{3.11}. First of all, let
us prove that problem \thetag{A.1} has a unique solution, which is smooth, in a
sufficiently small neighborhood of $(p_0,q_0)$. To this end, let us calculate
the first and the second derivatives of the function $D_\mu(p,q)$. We have
(denoting the derivatives by subscripts, omitting the arguments, and denoting
$D_\mu$ simply by $D$ and $F_\mu$ by $F$):
\begin{align}
D &= F_2 +\frac{\mu}{2} \|F_{1q}\|^2 + \frac{\mu}{2}\|F_{2q}\|^2,
\nonumber\\
D_q&= F_{2q}+\mu F_{1qq} F_{1q}+\mu F_{2qq} F_{2q}
= (I+\mu{\mathcal{E}}_2)F_{2q} + \mu {\mathcal{E}}_1 F_{2q},
\tag{A.3}\\
D_{qq} &= F_{2qq} + \mu F_{1qq} F_{1qq} +\mu F_{2qq} F_{2qq} +\cdots =
{\mathcal{E}}_2 +\mu{\mathcal{E}}_1^2 +\mu{\mathcal{E}}_2^2 +\cdots,
\nonumber
\end{align}
where $I$ is the identity matrix and the dots stand for terms linear in
$F_{1q}$ and $F_{2q}$. At the point $(p_0,q_0)$ we have
$F_{1q}=F_{2q}=D_q=0$, and the matrix $D_{qq}= {\mathcal{E}}_2 +\mu{\mathcal{E}}_1^2
+\mu{\mathcal{E}}_2^2$ is positive definite. Indeed, ${\mathcal{E}}_1$ and
${\mathcal{E}}_2$ are symmetric and ${\mathcal{E}}_2$ is positive semidefinite, and so
for any vector $\xi\in\R^n$ we have
\begin{align*}
\langle \xi, D_{qq}\xi\rangle &= \langle \xi,{\mathcal{E}}_2\xi\rangle +\mu
\langle {\mathcal{E}}_1\xi,{\mathcal{E}}_1\xi\rangle + \mu
\langle {\mathcal{E}}_2\xi,{\mathcal{E}}_2\xi\rangle\\
&= \langle \xi,{\mathcal{E}}_2\xi\rangle +\mu
\langle {\mathcal{E}}\xi,{\mathcal{E}}\xi\rangle\geqslant \mu\|\xi\|^2\|{\mathcal{E}}^{-1}\|^{-2}
\end{align*}
(here $(\cdot\,,\cdot)$ is the $\C^n$ inner product). It follows that
$D_{qq}$ is positive definite in a neighborhood of the point $(p_0,q_0)$,
and so the solution to problem \thetag{A.1} is unique and smooth and is
determined by the equation $D_q=0$. Set
$$
d(p)=d_\mu(p)=D(p,q(p)).
$$
Then, obviously,
$$
F_2(p,q(p)) \leqslant d(p)\quad\text{and}\quad \Big|\frac{\pa F_2}{\pa
q}(p,q(p))\Big| \leqslant \op{const} d(p)^{1/2}.
$$

It is quite obvious that, if we choose various $\mu>0$, then the resultant
functions $d_\mu(p)$ will be equivalent to each other; furthermore,
$q_\mu(p)-q_\nu(p)=O(d(p)^{1/2})$, and hence, writing $\De
q=q_\mu(p)-q_\nu(p)$, we have
\begin{align*}
{\wt F}_\mu(p)- {\wt F}_\nu(p) &= F_q sq +\frac12 \langle \De q, F_{qq}\De
q\rangle - F_q\De q -\frac12 \langle \De q, F_{qq}\De
q\rangle +O(\|F_q\|^3+\|\De q^3\|)\\
&= O(d(p)^{3/2}).
\end{align*}
Consequently, it suffices to prove the desired inequality for ${\wt F}_\mu(p)$
with $\mu$ arbitrarily small.

Let us calculate the imaginary part of the function $\wt F(p)={\wt F}_\mu(p)$
given by Eq.~\thetag{A.2}. In our shorthand notation, $\wt F(p) = \{F-1/2\langle
F_q,{\mathcal{E}}^{-1} F_q\rangle\}_{q=q(p)}$.

Let ${\mathcal{E}}^{-1}=A+iB$, where $A$ and $B$ are symmetric matrices with
real entries. Then
$$
(A+iB)({\mathcal{E}}_1+i{\mathcal{E}}_2)=I\quad (\text{the identity matrix}),
$$
and so
\begin{equation}
A{\mathcal{E}}_1 -B{\mathcal{E}}_2 = I,\qquad B{\mathcal{E}}_1 +A{\mathcal{E}}_2=0.
\tag{A.4}
\end{equation}
Let us prove that for any $\xi\in\R^n$
\begin{equation}
\langle \xi, B\xi\rangle\leqslant \ph\|\xi\|^2,
\tag{A.5}
\end{equation}
where $\ph=\ph(p,q)$ is a continuous function such that $\ph(p_0,q_0)=0$. Since
$$
\langle\xi, B\xi\rangle = \langle\xi, B(p_0,q_0)\xi\rangle +
\langle \xi,[B-B(p_0,q_0)]\xi\rangle,
$$
the estimate \thetag{A.5} will follow with $\ph(q)=\|B-B(p_0,q_0)\|$ if we
prove that $\langle\xi, B(p_0,q_0)\xi\rangle\leqslant 0$ for  any
$\xi\in\R^n$, or, equivalently $\Im\langle\xi,{\mathcal{E}}^{-1}\xi\rangle\leqslant0$
(we omit the argument $(p_0,q_0)$) for any
$\xi\in\R^n$. Let us take $\xi=i{\mathcal{E}}\eta$, $\eta=\eta_1+i\eta_2$,
$\eta_1,\eta_2\in\R^n$. Furthermore,
$$
\Im\langle\xi,{\mathcal{E}}^{-1}\xi\rangle = -\Im\langle\eta,{\mathcal{E}}\eta\rangle
= - \langle\eta_1,{\mathcal{E}}_2\eta_1\rangle +\langle\eta_2,{\mathcal{E}}_2\eta_2\rangle
- 2\langle\eta_2,{\mathcal{E}}_1\eta_1\rangle.
$$
Since $\xi\in\R^n$, it follows that ${\mathcal{E}}_1\eta_1={\mathcal{E}}_2\eta_2$,
and we obtain $\Im\langle\xi,{\mathcal{E}}^{-1}\xi\rangle
= -  \langle\eta_1,{\mathcal{E}}_2\eta_1\rangle
-\langle\eta_2,{\mathcal{E}}_2\eta_2\rangle\leqslant0$ (recall that ${\mathcal{E}}_2\geqslant 0$).

Let us now write
\begin{equation}
\wt F_2(p) = \Big\{F_2-\frac12\{\langle
F_{1q},BF_{1q}\rangle - \langle
F_{2q},BF_{2q}\rangle + 2\langle
F_{1q},AF_{2q}\rangle\}\Big\}\Big|_{q=q(p)}.
\tag{A.6}
\end{equation}
For $q=q(p)$ from \thetag{A.3} we have
\begin{equation}
F_{2q} = - (I+\mu{\mathcal{E}}_2)^{-1}\mu{\mathcal{E}}_1 F_{1q} = -\mu{\mathcal{E}}_1
F_{1q} +O(\mu^2) F_{1q},
\tag{A.7}
\end{equation}
where the $O(\mu^2)$ estimate is uniform whenever $q(p)$ lies in any given
bounded neighborhood of $q_0$ and $\|p-p_0\|<\e(\mu)$, where $\e(\mu)>0$ for
$\mu>0$.

Let us substitute \thetag{A.7} into \thetag{A.6} and take into account the
first equation in \thetag{A.4}. We obtain
\begin{align*}
\wt F_2 &= F_2-\frac12[\langle
F_{1q},BF_{1q}\rangle -2\mu \langle
F_{1q},(I+B{\mathcal{E}}_2)F_{1q}\rangle]+O(\mu^2)\\
&= F_2-\frac12 \langle
F_{1q},BF_{1q}\rangle +\mu\langle
F_{1q},F_{1q}\rangle -\mu\langle
BF_{1q},{\mathcal{E}}_2F_{1q}\rangle +\mathcal{F},
\end{align*}
where
\begin{equation}
|\mathcal{F}|\leqslant C\mu^2\|F_{1q}\|^2
\tag{A.8}
\end{equation}
whenever $q=q(p)$ lies in a sufficiently small neighborhood of $q_0$, and
$\|p-p_0\|\leqslant\e(\mu)$; the constant $C$ is independent of $\mu$ and
$p$. Set $\wt B(p,q)=B(p,q)-\ph(p,q)$. Then $\wt B\leqslant 0$ and
$$
\wt F_2 = F_2-\frac12\langle
F_{1q},\wt BF_{1q}\rangle +\Big(\mu-\frac\ph2\Big)\langle
F_{1q},F_{1q}\rangle
- \mu \langle \wt BF_{1q},{\mathcal{E}}_2F_{1q}\rangle -\mu\ph \langle
F_{1q},{\mathcal{E}}_2F_{1q}\rangle +\mathcal{F}.
$$
Since the matrix $\wt B$ is symmetric and nonpositive, we have
$$
\|\wt B F_{1q}\|^2 \leqslant -C_1 \langle
F_{1q},\wt B F_{1q}\rangle;
$$
furthermore, for any $\la>0$ we have
$$
|\langle \wt B
F_{1q},{\mathcal{E}}_2F_{1q}\rangle| \leqslant \frac12 \la\|\wt B F_{1q}\|^2
+\frac{\|{\mathcal{E}}_2\|^2}{\la}\|F_{1q}\|^2
\leqslant -\frac{C_1\la}{2} \langle
F_{1q},\wt B F_{1q}\rangle + \frac{\|{\mathcal{E}}_2\|^2}{\la}\|F_{1q}\|^2.
$$
By combining \thetag{A.8} with the subsequent two estimates, we obtain
\begin{equation}
\wt F_2 \leqslant F_2 -\frac12(1-C_1\la\mu)\langle F_{1q},\wt B F_{1q}
\rangle+ \Big[\mu-\frac{\ph}{2}-
\frac{\mu\|{\mathcal{E}}_2\|^2}{\la}-\mu\ph \|{\mathcal{E}}_2\| - C\mu^2\Big]\|F_{1q}\|^2.
\tag{A.9}
\end{equation}
Now set $\la=\ph\op{sup}\|{\mathcal{E}}_2\|^2$ and $\ph_0=1/\ph\sup\|\mathcal{E}_2\|^2$.
The coefficient of $\|F_{1q}\|^2$ in \thetag{A.9} will not be less
than $\mu/2-C\mu^2-\ph/2$. Set $\mu=\min\{1/(C_1\la), 1/(4C)\}$. Then the
coefficient of $\langle F_{1q}, \wt B F_{1q}\rangle$ in \thetag{A.9} will be
nonpositive, and the coefficient of $\|F_{1q}\|^2$ will be greater than or
equal to $\mu/4-\ph/2$. Now that $\mu$ is fixed, we can choose a neighborhood
of $p_0$ small enough to have $\ph(q(p))<\mu/4$ for $p$ in that
neighborhood. For these $p$ we obtain
$$
\wt F_2(p)\geqslant \{F_2+ c_3\|F_{1q}\|^2\}\big|_{q=q(p)} \geqslant \{F_2 +
c_4\|F_q\|^2\}\big|_{q=q(p)},\quad c_4>0,
$$
in view of \thetag{A.7}.

Thus,
$$
c_5\,dp \leqslant \wt F_2(p)\leqslant c_6\,d(p)
$$
in a small neighborhood of $p_0$ (the right inequality is obvious from
\thetag{A.2}).

The lemma is proved.

\section*{Acknowledgments}
The work was supported by the ISF under grant No. MFO000.

\end{document}